\documentclass[11pt]{article}
\usepackage[margin=1in,letterpaper]{geometry}
\usepackage[T1]{fontenc}
\usepackage[sc]{mathpazo}
\usepackage{array}
\usepackage{microtype}
\usepackage{amsmath}
\usepackage[dvipsnames]{xcolor}
\usepackage{amssymb}
\usepackage{amsthm}
\usepackage{bm,float}
\usepackage{dsfont}
\usepackage{authblk}
\usepackage{fullpage,caption,wrapfig}
\usepackage{comment}
\usepackage{mathtools}
\usepackage{threeparttable}
\usepackage{longtable}
\usepackage[shortlabels,inline]{enumitem}
\usepackage{bbm}
\usepackage{booktabs}
\usepackage[numbers]{natbib}
\usepackage{makecell}
\definecolor{pku-red}{RGB}{139,0,18}
\usepackage{nag,tikz}
\usetikzlibrary{calc}

\usepackage{tablefootnote}
\usetikzlibrary{decorations.pathreplacing}
\usetikzlibrary{graphs, positioning, quotes, shapes.geometric}
\usepackage{pgfplots}
\pgfplotsset{compat=1.18}
\usetikzlibrary{arrows.meta}%
\usetikzlibrary{arrows}
\usepackage{tikz-3dplot}
\usepgfplotslibrary{colormaps,fillbetween}

\usepackage{multicol}
\usepackage[scaled]{helvet} 

\usepackage{float}

\usepackage{multirow}

\usepackage[normalem]{ulem}

\usepackage{etoolbox}
\makeatletter
\patchcmd{\@bibitem}{\ignorespaces}{\label{bib-#1}\ignorespaces}{}{}
\makeatother

\usepackage[ruled]{algorithm2e} 

\SetAlFnt{\small}
\SetAlCapFnt{\small}
\SetAlCapNameFnt{\small}
\SetAlCapHSkip{0pt}
\IncMargin{-\parindent}

\usepackage{hyperref}
\usepackage{crossreftools}
\hypersetup{
    colorlinks=true,
    linkcolor=pku-red,     
    urlcolor=violet,
    citecolor=BlueViolet,
    pdffitwindow=true,
}\usepackage[capitalize, nameinlink]{cleveref}

\newcommand{\appref}[1]{\Cref{#1}}

\def\@cmidrulea{%
  \multispan\@cmidla&\multispan\@cmidlb
  \unskip\hskip\cmrkern@l%
  {\CT@arc@\leaders\hrule \@height\@thisrulewidth\hfill}%
  \hskip\cmrkern@r\cr
  \noalign{\nobreak\vskip-\@thisrulewidth} 
}%

\def\@cmidruleb{
  \multispan\@cmidlb
  \unskip\hskip \cmrkern@l%
  {\CT@arc@\leaders\hrule \@height\@thisrulewidth\hfill}%
  \hskip\cmrkern@r\cr
}%

\allowdisplaybreaks[3]

\theoremstyle{plain}
\newtheorem{theorem}{Theorem}[section]

\theoremstyle{definition}
\newtheorem{definition}[theorem]{Definition}
\newtheorem{remark}[theorem]{Remark}
\newtheorem{example}[theorem]{Example}

\DeclareMathOperator*{\supp}{supp}
\DeclareMathOperator*{\BR}{BR}

\newcommand{\R}{\mathbb{R}}

\newcommand{\poly}{\mathrm{poly}}

\newcommand{\T}{\mathsf{T}}

\title{A survey on algorithms for Nash equilibria in finite normal-form games}

\author[1]{Hanyu Li\thanks{\href{mailsto:lhydave@pku.edu.cn}{lhydave@pku.edu.cn}.}}

\author[2]{Wenhan Huang\thanks{\href{mailsto:wenhanhuang1993@gmail.com}{wenhanhuang1993@gmail.com}.}}

\author[1]{Zhijian Duan\thanks{\href{mailsto:zjduan@pku.edu.cn}{zjduan@pku.edu.cn}.}}

\author[3,5]{David Henry Mguni\thanks{\href{mailsto:d.mguni@qmul.ac.uk}{d.mguni@qmul.ac.uk}.}}

\author[3]{Kun Shao\thanks{\href{mailsto:shaokun2@huawei.com}{shaokun2@huawei.com}.}}

\author[4]{Jun Wang\thanks{\href{mailsto:jun.wang@cs.ucl.ac.uk}{jun.wang@cs.ucl.ac.uk}.}}

\author[1]{Xiaotie Deng\thanks{\href{mailsto:xiaotie@pku.edu.cn}{xiaotie@pku.edu.cn}.}}

\affil[1]{Peking University}
\affil[2]{Independent Researcher}
\affil[3]{Huawei Noah's Ark Lab}
\affil[4]{University College London}
\affil[5]{Queen Mary University London}

\date{\today}

\begin{document}
\hypersetup{pageanchor=false}
\maketitle 
\begin{abstract}
Nash equilibrium is one of the most influential solution concepts in game theory. With the development of computer science and artificial intelligence, there is an increasing demand on Nash equilibrium computation, especially for Internet economics and multi-agent learning. This paper reviews various algorithms computing the Nash equilibrium and its approximation solutions in finite normal-form games from both theoretical and empirical perspectives. For the theoretical part, we classify algorithms in the literature and present basic ideas on algorithm design and analysis. For the empirical part, we present a comprehensive comparison on the algorithms in the literature over different kinds of games. Based on these results, we provide practical suggestions on implementations and uses of these algorithms. Finally, we present a series of open problems from both theoretical and practical considerations.
\end{abstract}
\thispagestyle{empty} 

\newpage
\tableofcontents
\newpage
\setcounter{page}{1}
\hypersetup{pageanchor=true}
\pagenumbering{arabic}

\section{Introduction}

Game theory aims to formulate the interaction of players and figure out the outcome of the play. More specifically, people try to find different concepts of "equilibria" of the game, which are regarded as stationary game statuses. Among these equilibrium concepts, the most essential one is the \emph{Nash equilibrium}, named after John F. Nash, capturing a situation that no player can gain additional payoff by unilaterally deviating from the equilibrium strategy. Nash equilibrium can be used to analyze and explain hostile situations or cooperative outcomes in human behavior. For example, Nash equilibrium is used to explain behaviors in arms races~\cite{schelling1958strategy}. By predicting the occurrence of ``tragedy of the commons", it also propels regulation-making, e.g., versus industrial environmental pollution~\cite{van1992international}. With the upsurge of artificial intelligence research starting in the 21st century, Nash equilibrium guides machine learning methods towards finding equilibrium strategies of real-world competing games and consequently defeating several high-level human players, including AlphaStar for StarCraft II game~\cite{AlphaStar}, OpenAI Five for multiplayer online battle arena game~\cite{OpenAIFive}, and Suphx for Japanese mahjong game~\cite{Suphx}.

The concept of Nash equilibria can be traced back to the 1940s, when von Neumann and Morgenstern~\cite{von2007theory} first showed the existence of mixed Nash equilibria in two-player zero-sum games by proving the minimax theorem. Later, in the early 1950s, Nash extended the existence of such an equilibrium solution to any game with finite players and finite actions by employing Kakutani's fixed-point theorem \cite{nash1950equilibrium} and later using Brouwer's fixed-point theorem \cite{nash1951non}. Since then, non-cooperative games have embraced the Nash equilibrium as the most popular solution concept in constructing modern economic theories. Well-studied topics include auctions \cite{myerson1981optimal}, mechanism design \cite{myerson1989mechanism}, and the theory of the firm \cite{tirole1988theory}. The involvement of computer science brings even more demands and applications of Nash equilibria, including algorithmic game theory \cite{roughgarden2010algorithmic}, online learning \cite{sutton2018reinforcement}, and adversarial learning \cite{goodfellow2020generative}. Studies of these applications and game theory itself have developed a reciprocal relationship.

The original existence proof by Nash \cite{nash1950equilibrium,nash1951non} is non-constructive. From the view of computer science, an essential further concern is how to compute a Nash equilibrium. Unfortunately, a series of works \cite{DBLP:journals/jcss/Papadimitriou94,DBLP:journals/eccc/ECCC-TR05-115,DBLP:journals/siamcomp/DaskalakisGP09,DBLP:journals/eccc/ECCC-TR05-139,DBLP:journals/eccc/ECCC-TR05-134,DBLP:conf/focs/ChenD06,chen2009settling,papadimitriou_2007,smoothNash} showed that computing a Nash equilibrium can hardly admit any polynomial-time algorithm, even when we restrict to two-player games and allow random perturbation on the input. In fact, there are only very few (exponential-time) algorithms computing Nash equilibria in general two-player games \cite{Lemke-Howson}, and no algorithm exists for general games with more than two players to the best of our knowledge. Attention thus turns to find an approximation solution. In this survey, we mainly focus on two well-studied notions of approximation: $\epsilon$-approximate Nash equilibrium and $\epsilon$-well-supported Nash equilibrium. These notions are additive approximations and share intuitive economic interpretation: $\epsilon$ is the additional payoff gained by deviating under different meanings.

In the literature, exact and approximation algorithms for different kinds of games were proposed from various perspectives, including optimization, economics, learning, and the study of the approximation notion itself. Very recently, a number of works emerged, making progress on these fundamental approximation concepts in Nash equilibrium computation. On the other hand, empirical studies about Nash equilibrium approximation are rather rare even on two-player games, including only \cite{tsaknakis2008performance,DBLP:conf/wea/FearnleyIS15,DBLP:conf/wea/KontogiannisS11,chen_tightness_2023} to the best of our knowledge. The gap between theoretical results and practical performances becomes larger and larger. With the rapid growth of practical demands for Nash equilibrium computing (especially in artificial intelligence \cite{AlphaStar,mcmahan2003planning,lanctot2017unified}), we think it is the proper time to review representative methods in the literature and fill in a more complete empirical understanding of these methods. 

This survey is thus naturally divided into two parts. For the theoretical part, we review different notions of approximation and present a classification of algorithms in the literature. We demonstrate that how these algorithms are designed and analyzed based on a very few basic ideas. Such a classification also implies the pros and cons of different ideas, bringing a lot of open problems from new perspectives.

For the empirical part, based on variant performance criteria, we present a comprehensive comparison on the algorithms in the literature over different kinds of games. These criteria focus mostly on practical considerations, including actual approximation, efficiency, and precision requirements. Such comparison brings new phenomena that are not observed in the previous empirical studies \cite{DBLP:conf/wea/FearnleyIS15}. Our experiments also provide practical suggestions on implementations and uses of these algorithms. Moreover, new theoretical challenges are proposed from these new phenomena.

This paper is organized as follows. In \Cref{sec:definition}, we introduce the basic notions of Nash equilibria and review the complexity results on exact/approximate Nash equilibrium computation. In \Cref{sec:algorithm}, the existing algorithms for exact/approximate Nash equilibrium computation are reviewed and classified. In \Cref{sec:experiment}, we present the empirical comparison and the corresponding analysis of these algorithms. Finally, in \Cref{sec:conclusion}, we conclude this survey, provide suggestions on practical uses, and raise open problems from both theoretical and practical aspects.

\section{Computational aspects of Nash equilibria}
\label{sec:definition}

%
%

\subsection{Finite normal-form games}
We first introduce several concepts and notations. A finite \emph{normal-form game} consists of a finite number of players, each with a finite number of pure strategies. When all players choose pure strategies simultaneously, each will be given a certain number of payoff. In a game play, players are allowed to choose a mixed strategy, i.e., a probability distribution over pure strategies. Then, the goal of all players is to maximize their own expected payoff.

Since most algorithms in the literature only handle two-player games, we here introduce the formal notations on two-player games, or called \emph{bimatrix games} in the literature (including most literature mentioned in this survey). 

For a matrix $A$, denote the $i$th row by $A^i$ and the $j$th column by $A_j$. Notation $1_k$ represents a $k$-dimensional vector with all its entries being $1$. The standard orthonormal basis of $\R^k$ is denoted by $e^1_k,\ldots,e^k_k$. When there is no confusion about the dimension, we will simply write $e^i$ for the $i$th basis. The support of a vector $v$, denoted by $\supp(v)$, is the set of indices at which the entry of $v$ is nonzero. Denote the standard $(k-1)$-simplex by $\Delta_k$, which is the set $\{(x_1,\dots,x_k)\in\R^k:x_1+\dots+x_k=1,x_i\geq 0,\forall i\}$. 

A \textit{bimatrix game} has two players called the row player and the column player, each with $m$ and $n$ pure strategies. When the row player chooses the $i$th strategy and the column player chooses the $j$th strategy, their payoffs are $R_{ij}$ and $C_{ij}$, respectively. The payoffs thus form two $m\times n$ matrices $R$ and $C$. Two players are allowed to independently choose randomized strategies, which are called \textit{mixed strategies}. Their mixed strategies are distributed on $\Delta_m$ and $\Delta_n$, respectively. When two players choose (mixed) strategies $x$ and $y$ independently, we say $(x,y)$ is a \textit{strategy profile}. The row player's (expected) payoff is $x^\T Ry$ and the column player's is $x^\T C y$.

We summarize our notations in the following table.

\begin{table}[ht]
    \centering
    \caption{Summary of notations in this paper.}
    \label{tab:notation-summary}
    \begin{tabular}{cl}
    \toprule
         Notation & Description \\ \midrule\midrule
         $A^i$ & the $i$th row of matrix $A$\\
         $A_j$ & the $j$th column of matrix $A$\\
         $A_{ij}$ & the element at the $i$th row and the $j$th column of matrix $A$\\
         $\R^k$ & the $k$-dimensional Euclidean space\\
         $e_k^i$ (or $e^i$) & the $i$th vector of the standard orthonormal basis in $\R^k$\\
         $1_k$ & a $k$-dimensional vector with all its entries being $1$\\
         $\supp(v)$ & the support of vector $v$ \\
         $\Delta_k$ & the standard $(k-1)$-simplex\\
         $(R,C)$ & a bimatrix game with payoff matrices $R$ and $C$ \\
         $m$ & the number of row player's pure strategies\\
         $n$ & the number of column player's pure strategies\\
         $(x,y)$ & a strategy profile that row player chooses $x$ and column player chooses $y$\\\bottomrule
    \end{tabular}
\end{table}

\subsection{Approximation notions for Nash equilibria}

Following the literature~\cite{DBLP:conf/sigecom/LiptonMM03,kontogiannis2009polynomial}, we provide the following approximation notion.

\begin{definition}[$\epsilon$-approximate Nash equilibrium]
Let $(R, C)$ be a bimatrix game whose payoffs belong to $[0,1]$. For $\epsilon\geq 0$, a strategy profile $(x,y)$ is an $\epsilon$-approximate Nash equilibrium if
\begin{align*}
    {(e^i_m)}^\T Ry&\leq x^\T Ry+\epsilon,\quad i=1,\dots,m,\\
    x^\T Ce^j_n&\leq x^\T Cy+\epsilon,\quad j=1,\dots,n.
\end{align*}
\end{definition}

Less formally, a strategy profile is called an $\epsilon$-approximate Nash equilibrium if no player can gain more than $\epsilon$ payoff by deviating from the player's present strategy. The extra gain on payoff is called the \emph{incentive} \cite{bosse2010new,daskalakis2007progress} or \emph{regret} \cite{1/3-NE} of the player. Such a definition can be extended to multiplayer games in a natural way. Note that the payoff restriction in the definition can be viewed as the normalization of payoff matrices. It is also called \emph{$\epsilon$-Nash equilibrium} ($\epsilon$-NE) \cite{bosse2010new,kontogiannis2009polynomial}. In the literature, various names were raised for $\epsilon$: approximation \cite{bosse2010new,kontogiannis2009polynomial}, approximation ratio \cite{daskalakis2007progress}, approximation factor \cite{bosse2010new,DBLP:conf/sigecom/LiptonMM03}, approximation error \cite{bosse2010new}, or approximation guarantee \cite{DBLP:journals/algorithmica/CzumajDFFJS19}. In this paper, we simply call $\epsilon$ the \textit{approximation}.

A related concept in the machine learning context is \textit{exploitability} (termed by \cite{exploitability}), or the Nikaido-Isoda function~\cite{nikaido1955note}. Different from considering the maximum incentive, this concept sums up the incentives of all players. Notice that for bimatrix games, an $\epsilon$-approximate Nash equilibrium strategy profile has at most $2\epsilon$ exploitability. On the other hand, an $\epsilon$-exploitability strategy profile forms an $\epsilon$-NE. As these two concepts can be converted to each other, we focus on $\epsilon$-NE in this paper.

Another well-studied approximation concept is the \emph{$\epsilon$-well-supported Nash equilibrium} (WSNE). We follow \cite{smoothNash,DBLP:conf/sigecom/LiptonMM03} to present the following definition.

\begin{definition}[$\epsilon$-well-supported Nash equilibrium]
Let $(R, C)$ be a bimatrix game whose payoffs belong to $[0,1]$. A strategy profile $(x,y)$ is an $\epsilon$-well-supported Nash equilibrium ($\epsilon$-WSNE) if 
\begin{align*}
    \forall i\in\{1,\cdots,m\}\quad i\in\supp(x)&\implies R^jy\leq R^iy+\epsilon,j=1,\cdots,m,\\
    \forall i\in\{1,\cdots,n\}\quad i\in\supp(y)&\implies x^\T C_j\leq x^\T C_i+\epsilon,j=1,\cdots,n.
\end{align*}
\end{definition}
In our context, $\epsilon$ here is called the \textit{well-supported approximation} (similar to approximation).

WSNE is a stronger approximation notion than approximate Nash equilibria: If $(x,y)$ is an $\epsilon$-WSNE, then $(x,y)$ is an $\epsilon$-NE; however, the reverse is not true. Indeed, a small difference on strategies leads to a small difference in the approximation; however, this is not true for the well-supported approximation. The following example shows this fact.
\begin{example}\label{ex:WSNE-diff}
    Consider the following bimatrix game:
    \[R=\begin{pmatrix}
        1/3&0\\1&1
    \end{pmatrix},\qquad C=\begin{pmatrix}
        1/3&1\\0&1
    \end{pmatrix}.\]
    A Nash equilibrium is $x^*=y^*=(0,1)^\T$. Consider strategy profile $x'=x^*$ and $y'=(\epsilon,1-\epsilon)^\T$, where $\epsilon\in(0,1)$. $(x',y')$ is an $\epsilon$-NE. However, $(x',y')$ is a $1$-WSNE where $1$ is the minimum possible choice.
\end{example}

In 2009, Chen, Deng, and Teng \cite{chen2009settling} showed the reverse relationship: Given a $\epsilon^2/8$-NE, we can compute an $\epsilon$-WSNE in polynomial time. Thus, all complexity and approximation results stated for approximate Nash equilibria can be translated into the version for WSNE. 

\begin{remark}
In the literature, people also consider \emph{continuous games}, where each player has a continuous action space. A solution concept named \emph{local Nash equilibrium} characterizes a pure strategy profile in which no player can obtain more utility when deviating over a neighborhood of its strategy (an open set containing its strategy).

Several gradient-based optimizing methods are proposed to search this concept iteratively, e.g., extragradient (EG)~\cite{korpelevich1976extragradient}, optimistic gradient (OG)~\cite{daskalakis2017training}, mirror descent (MD)~\cite{nemirovskij1983problem}, consensus optimization (CO)~\cite{consensus-optimization}, gradient descent ascent (GDA)~\cite{lin2020gradient}, and Hamiltonian gradient descent (HGD)~\cite{mescheder2017numerics, balduzzi2018mechanics}.

It should be noticed that local Nash equilibria in continuous games differ from Nash equilibria in bimatrix games. Specifically, consider a zero-sum bilinear continuous game with payoff $(x^\T Ry, -x^\T Ry)$. Viewing $x,y$ as continuous strategies, the $0$-vector is a trivial local Nash equilibrium. However, viewing $x,y$ as mixed strategies in a zero-sum bimatrix game, the $0$-vector contributes nothing to finding the Nash equilibrium. Although some methods can be applied to bimatrix games and find coarse correlated equilibrium, e.g., optimistic mirror descent with smooth regularization~\cite{anagnostides2022optimistic}, we do not consider local Nash equilibrium nor continuous games in this paper.
\end{remark}

\subsection[Lower-bound and upper-bound results]{Lower-bound and upper-bound results of approximation notions on Nash equilibria}\label{subsec:complexity-NE}

In computational complexity, lower-bound results show that the problem is hard to solve to some extent. Usually, the hardness is established by showing that this problem is $X$-hard for some complexity class $X$ or that this problem requires a certain amount of time to provide an answer. On the other hand, upper-bound results means that the problem admits an algorithm with certain approximation or complexity. Lower bounds and upper bounds together capture the computational nature of this problem. 

As for NE, an overall picture is that it hardly admits a polynomial-time algorithm even when
\begin{enumerate*}
    \item restricting to two-player games,
    \item allowing a small random perturbation on the input, and
    \item only seeking for a small constant approximation.
\end{enumerate*}
On the other hand, however, the average cases have polynomial-time algorithms and thus are much easier to solve. For an introduction on proof techniques, see \cite{CHEN200788} and \cite{DASKALAKIS200987}.

To make a more formal presentation, one should first note that it is more appropriate to consider computing $2^{-n}$-NE, rather than exact NE (see \cite{etessamiComplexityNashEquilibria2010} and \cite{gargRCompletenessDecisionVersions2018} for further discussions), which is called \textsc{Nash} for general games and $r$-\textsc{Nash} for $r$-player games in the literature \cite{DBLP:journals/eccc/ECCC-TR05-115, DBLP:journals/eccc/ECCC-TR05-139, DBLP:journals/eccc/ECCC-TR05-134, DBLP:conf/focs/ChenD06, chen2009settling}. ($r$-)\textsc{Nash} is proved to be a natural problem that lies in $\mathrm{PPAD}$ \cite{DBLP:journals/eccc/ECCC-TR05-115}, a complexity class introduced by Papadimitriou \cite{DBLP:journals/jcss/Papadimitriou94} in 1994. Later, finding a Nash equilibrium was shown to be $\mathrm{PPAD}$-hard in games with at least four players by Daskalakis, Goldberg, and Papadimitriou \cite{DBLP:journals/eccc/ECCC-TR05-115}; independently for three players by the two teams~\cite{DBLP:journals/eccc/ECCC-TR05-139, DBLP:journals/eccc/ECCC-TR05-134}; and finally in bimatrix games by Chen and Deng \cite{DBLP:conf/focs/ChenD06}. Moreover, computing $1/\poly(n)$-NE is also shown to be $\mathrm{PPAD}$-hard \cite{DBLP:journals/siamcomp/DaskalakisGP09, chen2009settling}. It is well-believed that $\mathrm{PPAD}$-hard problems admit no polynomial-time algorithms \cite{papadimitriou_2007}, leading to the hardness of Nash equilibrium computing. Chen, Deng, and Teng~\cite{smoothNash} further showed that under a well-accepted assumption ($\mathrm{PPAD}\not\subseteq \mathrm{RP}$), no algorithm computing a Nash equilibrium in a two-player game algorithm runs in polynomial time even when the input is randomly perturbed. Note that by the above discussion, computing $\epsilon$-NE and $\epsilon$-WSNE makes no difference when $\epsilon$ is $2^{-n}$ or $1/\poly(n)$. Thus, the above results also hold for a well-supported approximation.

Due to the above hardness results, it is more reasonable to consider $\epsilon$-NE and $\epsilon$-WSNE when $\epsilon$ is a \emph{constant}. Lipton, Markakis, and Mehta \cite{DBLP:conf/sigecom/LiptonMM03} showed that $\epsilon$-NE can be found by brute-force search in quasi-polynomial time\footnote{It means the running time is $2^{O\poly(n)}$. Quasi-polynomial time is slower than polynomial time, but faster than exponential time. } for arbitrary constant $\epsilon>0$. More precisely, they showed that one can find an $\epsilon$-NE by an exhaustive search on a pair of $k$-uniform strategies\footnote{A mixed strategy is called $k$-uniform if it is the uniform distribution on a multiset $S$ of pure strategies, with $|S| = k$.} with $k=\lceil12\log n/\epsilon^2\rceil$. An similar result of reaching $\epsilon$-WSNE by a $k$-uniform strategy profiles with $k=\lceil 2\log(2n)/\epsilon^2\rceil$ was also proved by Kontogiannis and Spirakis \cite{kontogiannis2007efficient} in 2007. Evidence indicates that such quasi-polynomial results appear tight (up to the $o(1)$ term): Rubinstein \cite{DBLP:conf/focs/Rubinstein16} showed that there exists a constant $\epsilon^*>0$ such that, assuming Exponential Time Hypothesis for $\mathrm{PPAD}$\footnote{Ruoghly speaking, this means that solving a $\mathrm{PPAD}$-complete problem like $r$-\textsc{Nash} requires an exponential time.}, computing an $\epsilon^*$-NE in two-player games requires time $n^{\log^{1-o(1)}n}$. However, $\epsilon$-NE can be computed in polynomial time if we consider special kinds of games. For example, if the number of non-zero entries in any column of $R+C$ in a bimatrix game is at most $s$, then $\epsilon$-NE can be computed in time $n^{O(\log s/\epsilon^2)}$ \cite{barmanApproximatingNashEquilibria2015}.

From another direction, it is natural to ask for the minimum $\epsilon$ that admits a polynomial-time algorithm. For $\epsilon$-NE, the state-of-the-art $\epsilon$ is $1/3 + \delta$ ($\delta$ is an arbitrary constant), given by Deligkas, Fasoulakis, and Markakis \cite{1/3-NE} in 2022, which revises the previous $(0.3393 + \delta)$-approximate algorithm by Tsaknakis and Spirakis \cite{tsaknakis2008optimization} in 2007. For $\epsilon$-WSNE, the state-of-the-art $\epsilon$ is $1/2 + \delta$ ($\delta$ is an arbitrary constant), given by Deligkas, Fasoulakis, and Markakis \cite{1/2-WSNE} in 2022.

The complexity results are also established for other approximation notions. If we consider relative approximation rather than the above additive approximations, then Daskalakis \cite{daskalakisComplexityApproximatingNash2013} established the $\mathrm{PPAD}$-hard result for two-player games and constant approximation. When considering additive approximation but the payoff is "normalized" to $[-1,1]$ rather than $[0,1]$, computing $\epsilon$-NE and $\epsilon$-WSNE is $\mathrm{PPAD}$-hard for constant $\epsilon$ even when the input is allowed to have a constant magnitude of perturbation \cite{DBLP:conf/focs/BoodaghiansBHR20}.

If we require a reasonable property on Nash equilibria, such as playing a certain pure strategy with positive probability, then deciding the existence of such Nash equilibria becomes $\mathrm{NP}$-hard, even in two-player games \cite{newComConitzer08}. Since $\mathrm{PPAD}\subseteq\mathrm{NP}$, the above result indicates that computing NE could become even harder when more restrictions are provided on NE.

From all the above results, it seems that Nash equilibrium and even its approximation notions are hard to find in polynomial time, even in two-player games. However, when we consider average cases, i.e., assuming the input game follows a certain probability distribution (also called \emph{random games} \cite{baranyNashEquilibriaRandom2007, nudelman2004run, panagopoulouRandomBimatrixGames2011}), Nash equilibria are much easier to find than in general cases. Indeed, in two-player games, if every entry in each of the matrices is drawn independently according to the uniform distribution on $[0,1]$ or the standard Normal distribution $\mathcal{N}(0,1)$, then there exists an algorithm finding Nash equilibria, whose running time is polynomial with high probability \cite{baranyNashEquilibriaRandom2007}. In addition, if the number $n$ of strategies in a two-player random game tends to infinity, the uniform strategy profile is a $\sqrt{\log n/n}$-NE and a $\sqrt{3\log n/n}$-WSNE with high probability \cite{panagopoulouRandomBimatrixGames2011}; in words, random bimatrix games are asymptotically easy to solve.
\section{Theoretical classification of algorithms for Nash
equilibrium computation}\label{sec:algorithm}

Now we review all existent algorithms in the literature to our knowledge for Nash equilibrium computation in finite normal-form games. Algorithms can be divided into following classes. Since the existing algorithms mostly focus on two-player games, we separately consider those working for general $r$-player games with $r\geq 3$. 
\begin{enumerate}
    \item Brute-force search.
    \item Algorithms designed to calculate an exact NE. The running time may not be a polynomial.
    \item Polynomial-time algorithms designed to calculate an $\epsilon$-NE or an $\epsilon$-WSNE with a guaranteed fixed real number $\epsilon$.
    \item Heuristic algorithms with limited theoretical analysis on convergence and approximation, mostly learning dynamics. They also work for multi-player games.
    \item Algorithms working for games of more than two-players with theoretical guarantees.
\end{enumerate}

\subsection{Brute-force search}\label{subsec:brute-force}

We begin with brute-force search algorithms. Usually, brute-force search is trivial for a computational problem and not necessary to mention, but it is not the case for NE. Even more surprisingly, such algorithms only exist for two-player games when we consider exact NE. In fact, there is no known algorithm computing exact NE for general games with more than two players. For two-player games, brute-force search relies on the indifference principle: A Nash equilibrium should be a strategy profile such that for each player, all pure strategies played with positive probability attain the same payoff. Thus, by enumerating all possible supports in a strategy profile, we can calculate Nash equilibria by solving linear equations. However, the brute-force method usually takes exponential time and thus is inefficient.

For $\epsilon$-NE, a similar brute-force method was discovered by Lipton, Markakis, and Mehta \cite{DBLP:conf/sigecom/LiptonMM03}. They provide a quasi-polynomial-time algorithm by enumerating $k$-uniform strategies with $k=\lceil12\log n/\epsilon^2\rceil$ (see \Cref{subsec:complexity-NE}). Such enumerations on strategies cannot be improved even in two-player win-lose games\footnote{That is, all entries in $R,C$ are $0,1$ and all entries in $R+C$ are $1$.} \cite{feder2007approximating}. As for approximate WSNE, due to the similar result by Kontogiannis and Spirakis \cite{kontogiannis2007efficient}, an exhaustive search for $k$-uniform strategy profiles with $k=\lceil 2\log(2n)/\epsilon^2\rceil$ will find an $\epsilon$-WSNE. The total running time for this method is again quasi-polynomial.

The above brute-force search algorithms do not run in polynomial time, which matches the lower-bound results in \Cref{subsec:complexity-NE}.

\subsection[Computing exact Nash equilibria in two-player zero-sum games]{Computing exact Nash equilibria in two-player zero-sum games: linear programming}
The earliest success in the design of algorithms computing exact Nash equilibria comes from zero-sum (or constant-sum) two-player games, whose payoff matrices satisfy $R+C=0$ (or all entries in $R+C$ are the same value). Such algorithms rely on a special property shared only by two-player zero-sum games: von Neumann's minimax theorem \cite{neumann1928theorie}, i.e.,
\[\max_{x\in\Delta_m}\min_{y\in\Delta_n}x^\T Ry=\min_{y\in\Delta_n}\max_{x\in\Delta_m}x^\T Ry=-\max_{y\in\Delta_n}\min_{x\in\Delta_m} x^\T Cy.\]
This result implies that a Nash equilibrium is reached when each player maximizes their own utility in the worst cases against the opponent. One popular method utilizing this theorem is solving Nash equilibria by linear programs, which have polynomial-time algorithms such as ellipsoid methods \cite{shorConvergenceRateGradient1970} and interior methods \cite{karmarkarNewPolynomialtimeAlgorithm1984}.

Specifically, optimal solution pair $(x^*,y^*)$ of the following primal and dual linear programs are precisely Nash equilibria with payoff matrices $R$ and $C=-R$:
\begin{center}
\begin{multicols}{2}

Primal LP (with variables $x\in\R^m$ and $t\in\R$):
\begin{align*}
    \mathrm{maximize\quad} &t\\
    \mathrm{subject~to\quad} &R^\T x\geq t1_n,\\
    &x\geq 0,\\
    &1_m^\T  x=1.
\end{align*}

Dual LP (with variables $y\in\R^n$ and $w\in\R$):
\begin{align*}
    \mathrm{minimize\quad} &w\\
    \mathrm{subject~to\quad} &Ry\leq w1_m,\\
    &y\geq 0,\\
    &1_n^\T  y=1.
\end{align*}
\end{multicols}
\end{center}

\subsection[Computing exact Nash equilibria in general two-player games]{Computing exact Nash equilibria in general two-player games: Lemke-Howson algorithm and linear complementarity problems}\label{subsec:LH-LCP}

To compute an optimal solution for a linear program, the simplex method was developed by the pioneering work of Dantzig in 1947 \cite{DANTZIG198243}. It is widely used, although it may take exponential time. The simplex method utilizes the geometric structure of the space of feasible solutions and searches for an optimal solution by traversing vertices. In 1964, a decade or so after Nash’s seminal work, following the idea of the simplex method, Lemke and Howson \cite{Lemke-Howson} developed a path-following algorithm for finding Nash equilibria in general two-player games.

Although the Lemke-Howson algorithm is proven to find a Nash equilibrium, analogous to the simplex algorithm, Savani and von Stengel \cite{savaniExponentiallyManySteps2004} showed that the Lemke-Howson algorithm could take exponential time. Moreover, the Lemke-Howson algorithm only works for so-called \emph{degenerate games} (see below). To solve these problems, heuristic improvements were proposed \cite{codenottiExperimentalAnalysisLemkeHowson2008, gattiCombiningLocalSearch2012}.

In fact, the Lemke-Howson algorithm appears in a more general context: to solve linear complementarity problems (LCP) \cite{LCP}. Just as viewing Nash equilibrium computing for zero-sum two-player games as linear programming, for general two-player games, LCP provides another perspective for computing Nash equilibria.

LCP aims to find a vector in real Euclidean space that satisfies a certain system of inequalities. Given a vector $q\in\R^n$, a $n\times m$ matrix $M$, LCP is to find a vector $z$ such that
\begin{align*}
    z&\geq 0,\\
    w=q+Mz&\geq 0,\\
    z^\T w&=0,
\end{align*}
or to show the nonexistence of such vectors. LCP can be viewed as a unifying of linear programming, quadratic programming, and Nash equilibrium computing in bimatrix games: All three kinds of problems can be written in the form of LCP. One can refer to \cite{LCP} for a more complete introduction.

Now we show how to make such a conversion on Nash equilibrium computing. Similar to the case of the Lemke-Howson algorithm, without loss of generality, we assume that all entries of $R$ and $C$ are positive. Consider the following LCP:
\begin{align*}
    x\geq 0,\quad u&=-1_m+Ry\geq 0,\quad x^\T u=0,\\
    y\geq 0,\quad v&=-1_n+C^\T x\geq 0,\quad y^\T v=0.
\end{align*}
Let $(x', y')$ be a solution for this LCP. Then it can be checked that the normalization $x^*=x'/(1_m^\T x')$ and  $y^*=y'/(1_n^\T y')$ form a Nash equilibrium. Conversely, if $(x^*,y^*)$ is a Nash equilibrium, then pair $x'=x^*/((x^*)^\T Cy^*)$, $y'=y^*/((x^*)^\T Ry^*)$ is a solution of the LCP. Thus solutions of this LCP exactly correspond to Nash equilibria of game $(R, C)$.

Finally, we begin to state the Lemke-Howson algorithm. Readers may skip this part if they do not want to get into details. Without loss of generality, suppose that all entries of $R$ and $C$ are nonnegative. We further assume that $R$ has no all-zero columns and $C$ has no all-zero rows. All of these assumptions can be satisfied by adding a sufficiently large positive number to all entries of $R$ and $C$, which does not affect the structure of Nash equilibria. Define two polytopes 
\begin{align*}
    P_1&=\left\{x\in\R^m: x_i\geq 0,i=1,\dots,m,x^\T C^j\leq 1,j=1,\dots,n\right\},\\
    P_2&=\left\{y\in\R^n: y_i\geq 0,i=1,\dots,n,R_jy\leq 1,j=1,\dots,m\right\}.
\end{align*}
We \emph{do not} require that $x$ and $y$ be normalized (probability distributions). The normalization of $x$ (if it exists) is denoted by $x^*$. Two polytopes characterize the support and best response of a strategy profile. Specifically, let $x\in P_1$. Observe the following properties:
\begin{enumerate}
    \item If $x_i=0$, then the strategy $x^*$ does not support on row $i$. 
    \item If $x^\T C_j=1$, then column $j$ attains the maximum possible payoff among all pure strategies, i.e., $j$ is a best response column.
\end{enumerate}
To make a concise statement over these properties, we introduce the notion of \emph{labels}. Let the pure strategy set of the row player be $M=\{1,\dots,m\}$ and the one of the column player be $N=\{m+1,\dots,m+n\}$. We say $x\in P_1$ \emph{has a label $k\in M\cup N$} if either $k\in M$ and $x_k=0$ or $k\in N$ and $x^\T C_k=1$. Labels are a key concept to Nash equilibria: It is not hard to show that nonzero $x\in P_1$ and nonzero $y\in P_2$ yields a Nash equilibrium $(x^*,y^*)$ if and only if $x$ and $y$ together have all labels from $1$ to $m+n$.

We need one more important concept before describing the Lemke-Howson algorithm: \emph{non-degenerate games}. We say a polytope $P$ is ($d$-dimension) \emph{simple} if at every vertex of $P$, there hold exactly $d$ equalities among the inequalities forming this vertex. A game is \emph{non-degenerate} if polytopes $P_1$ and $P_2$ are simple. Similar to situations in linear programming, a slight perturbation will make $P_1$ and $P_2$ simple and thus make a game non-degenerate. So perturbations help the Lemke-Howson algorithm to find $\epsilon$-NE with arbitrarily small $\epsilon$ in degenerate games.

The Lemke-Howson algorithm starts from $(x,y) = (0,0)$ and moves from one vertex to another vertex\footnote{This step is called \emph{pivoting}, just the same as the simplex algorithm.}. Note that the starting point has all labels. Since every label is a representative of an inequality, moving from one vertex to another vertex is the same as removing a label and adding a new label. The Lemke-Howson algorithm will take any label $k_0$ at the beginning. Once $k_0$ is added after pivoting, then it stops.

Analysis shows that during pivoting, labels in $(M\cup N)\setminus\{k_0\}$ are always held by either $x$ or $y$. Moreover, vertices visited by pivoting have a degree of one ($x$ and $y$ have all labels) or two ($x$ and $y$ share a duplicate label). Thus the Lemke-Howson algorithm can follow either a path or a cycle. Another fact mentioned previously is that every vertex with degree one (i.e., the endpoint of a path, or $x$ and $y$ have all labels) except $(0,0)$ is a Nash equilibrium. Since the Lemke-Howson algorithm starts at $(0,0)$, whose degree is one, a Nash equilibrium must be reached by following a path.\footnote{Clearly, there are an even number of endpoints of all paths. A corollary of this result is that the number of Nash equilibria in a non-degenerate game is odd.} Thus, the algorithm is guaranteed to find a Nash equilibrium in a finite number of iterations in non-degenerate games.

\subsection{Finding \texorpdfstring{$\epsilon$}{epsilon}-NE in two-player games: search-and-mix methods}\label{subsec:epsNE}

Now we turn to approximation algorithms with theoretical guarantees. Works on improving the constant approximation arose and thrived in 2006, which was exactly the time when $2$-\textsc{Nash} was proved to be $\mathrm{PPAD}$-complete. The approximation was consistently improved from $3/4$ to $0.3393$ in 2006 and 2007 \cite{kontogiannis2009polynomial, daskalakis2009note, daskalakis2007progress, bosse2010new, tsaknakis2008optimization}. The state-of-the-art algorithm with a $0.3393$ approximation was then given by Tsaknakis and Spirakis (TS algorithm) \cite{tsaknakis2008optimization}. Since then, however, no new algorithms securing a better approximation have been devised for 15 years. In 2022, Deligkas, Fasoulakis, and Markakis \cite{1/3-NE} improved the mixing method of the TS algorithm and finally got an approximation of $1/3$. In the literature, all polynomial-time algorithms with a theoretical approximation guarantee can be described as a two-phase method: Search for polynomial-time solvable solution concepts (search phase), and then make a convex combination over these solutions (mixing phase), which was firstly observed in \cite{chen_tightness_2023} and formalized in \cite{dengSearchandMixParadigmApproximate2023}.

Now we introduce the representative algorithms. For brevity, we name the algorithm in the form of author initials plus the publishing year and the guaranteed approximation, for example, TS07-0.3393. A summary of these polynomial-time algorithms is presented in \Cref{tab:epsNE}.
 
\begin{table}[ht]
    \centering
    \caption{Summary of polynomial-time algorithms computing $\epsilon$-NE with constant $\epsilon$. Many works have a journal version. Our citations here adopt the journal version, if possible. The approximation $x+\delta$ means that given $\delta>0$, one can calculate a $(x+\delta)$-NE in polynomial time of game size.}
    \label{tab:epsNE}
    \begin{tabular}{ccc}
    \toprule
         Author initials & Publish year & Approximation \\\midrule\midrule
         KPS \cite{kontogiannis2009polynomial} & 2006 & $0.75$ \\
         DMP \cite{daskalakis2009note} & 2006 & $0.5$ \\
         DMP \cite{daskalakis2007progress} & 2007 & $0.38+\delta$ \\
         BBM \cite{bosse2010new} & 2007 & $0.36$ \\
         TS \cite{tsaknakis2008optimization} & 2007 & $0.3393+\delta$ \\
         CDFFJS \cite{DBLP:journals/algorithmica/CzumajDFFJS19} & 2015 & $0.38$ \\
         DFM \cite{1/3-NE} & 2022 & $1/3+\delta$ \\\bottomrule
    \end{tabular}
\end{table}

\subsubsection*{KPS06-0.75}

The search phase in this algorithm finds two strategy profiles that reach the maximum payoff of each player respectively. Formally, it finds $i_1$, $j_1$, $i_2$, $j_2$ such that $R_{i_1,j_1}=\max_{i,j} R_{ij}$ and $C_{i_2,j_2}=\max_{ij} C_{ij}$.

The mixing phase simply mixes these pure strategies equally, i.e., to generate $\hat{x}=(e_m^{i_1}+e_m^{i_2})/2$ and $\hat{y}=(e_n^{j_1}+e_n^{j_2})/2$.

A direct calculation shows that $(\hat{x},\hat{y})$ is a $3/4$-NE.

\subsubsection*{DMP06-0.5}

The search phase in this algorithm is as follows. It first randomly picks a pure strategy of the row player, say row $i$. Then it finds the best response of the column player against row $i$, say column $j$. Similarly, it then finds the best response of the row player against column $j$, say row $i'$.

The mixing phase simply mixes row $i$ with $i'$ equally and generates $x^*=(e_m^i+e_m^{i'})/2$, $y^*=e_n^j$.

By calculations similar to KPS06-0.75, it can be shown that $(x^*,y^*)$ is a $1/2$-NE.

It is worth noting that the first two algorithms only involve \emph{two} pure strategies for each player. \cite{DBLP:conf/sigecom/LiptonMM03} showed that one can reach an $\epsilon$ approximation with support size $O(\log n/\epsilon^2)$. A natural question thus raises itself here: What is the best approximation one can get using a \emph{constant} number of pure strategies? The answer is $1/2$, given by Feder, Nazerzadeh, and Saberi \cite{feder2007approximating}. Their work showed that for any $\epsilon\in[0,1]$ and large integer $k$, there exists a bimatrix game such that the approximations of strategy profiles supported on $k$ rows and an arbitrary number of columns are no less than $1/2-\epsilon$. Thus BMP06-$1/2$ is the best algorithm one can expect when a constant number of pure strategies are used.

The following three algorithms below implement the search phase with linear programs (or solving zero-sum games). LPs make it possible to find strategies with larger supports and thus gain an approximation lower than $1/2$.

\subsubsection*{CDFFJS15-0.38}

\cite{DBLP:journals/algorithmica/CzumajDFFJS19} gave a distributed method to compute NE and WSNE. The analysis of the mixing phase of this method can be regarded as a standard technique for approximation analysis for $\epsilon$-NE algorithms. Thus, we present the detailed idea for the design and analysis of this algorithm.

In the search phase, the algorithm solves a Nash equilibrium $(x^, y^)$ for the zero-sum game $(R, -R)$ and a Nash equilibrium $(\hat{x}, \hat{y})$ for the zero-sum game $(-C, C)$, respectively. The value secured by $x^*$ in $(R,-R)$ is $v_r$. Similarly, the value secured by $\hat{y}$ in $(-C,C)$ is $v_c$. Note that in this phase, two players can do these calculations independently.

In the mixing phase, both payoff matrices are involved. The first idea is to combine solutions obtained in the search phase. Suppose, without loss of generality, that $v_r\geq v_c$. In the game $(R, C)$, the maximum payoff that the row player can achieve against $y^*$ is $v_r$. Thus, when the row player plays $\hat{x}$, the regret is at most $v_r$. Similarly, the regret of the column player playing $y^*$ in the game $(R, C)$ is at most $v_c$. Now, $(\hat{x},y^*)$ is a $v_r$-NE. However, there is no restriction on $v_r$. We must find another strategy profile to create a new restriction on the approximation. 

Below, our discussion on best responses is based on game $(R, C)$. Let column $j$ be a best response for the column player against $x^*$. Let row $r$ be a best response for the row player against column $j$. Now we make convex combinations over these mentioned strategies. To make the analysis tractable, we choose $(x',e_n^j)$ with $x' = p x^* + (1 - p)e_m^r$. The next step is to calculate the regret of this strategy profile for each player. The incentive of the row player is bounded by
\begin{align*}
    R_{rj} - (x')^\T Re^j_n &= R_{rj} - p(x^*)^\T Re^j_n - (1 - p)(e_m^r)^\T Re^j_n \\
    &= pR_{rj} - p(x^*)^\T Re^j_n \qquad (\text{since $(e_m^r)^\T Re^j_n = R_{rj}$}) \\
    &\leq p \cdot 1 - p(x^*)^\T Re^j_n \qquad (\text{since $R_{rj} \leq 1$}) \\
    &\leq p - p \cdot v_r \qquad (\text{applying minimax theorem on $(x^*,y^*)$}) \\
    &= p(1 - v_r).
\end{align*}
Similarly, the regret of the column player is bounded by $1 - p$.

By choosing the one with the minimum approximation between $(x',e_n^j)$ and $(\hat{x},y^*)$, the approximation of the output then has an upper bound $\min\left\{v_r, p(1 - v_r), 1 - p\right\}$. Choose $p$ so that $p(1 - v_r) = 1 - p$, i.e., $p = 1/(2 - v_r)$. Then the approximation can be bounded by $\min\{v_r, (1 - v_r)/(2 - v_r)\}$, denoted by $g(v_r)$. Note that $v_r \in [0, 1]$. Some calculations show that $g(v_r) \leq (3 - \sqrt{5})/2 \approx 0.382$. Thus, such a method finds a $0.382$-NE.

Now we summarize the procedure of design and analyzing the mixing phase of CDFFJS15-0.38. It first tries solutions in the search phase, which gives an approximation expressed by some constrained variables. However, such an expression is possibly not desirable. Then it sets undetermined coefficients for convex combinations over the solutions and the best responses. The corresponding strategy profile provides another approximation expressed by some constrained variables. Combining these two strategy profiles yields an approximation bounded by a constant real number.

\subsubsection*{BBM07-0.36}

Similar to CDFFJS15-0.38, BBM07-0.36 implements the search phase by finding a Nash equilibrium $(x^*,y^*)$ of zero-sum game $(R-C,C-R)$. Then we return to the original game $(R, C)$. Suppose $g_1$ is the regret of the row player playing $x^*$ against $y^*$ and $g_2$ is the regret of the column player playing $y^*$ against $x^*$. 

In the mixing phase, the algorithm also finds best responses. Suppose, without loss of generality, that $g_1\geq g_2$. Suppose row $r$ is the best response against $y^*$. Then the row player plays a mixture $\hat{x}=\delta_1 e_m^r+(1-\delta_1)x^*$, where $\delta_1$ is a function of $g_1$. Let column $b$ be the best response against $\hat{x}$. Suppose $h=(x^*)^\T Ce_n^b-(x^*)^\T Cy^*$. The column player plays a mixture $\hat{y}=\delta_2 e_n^b+(1-\delta_2) y^*$, where $\delta_2$ is a function of $\delta_1,g_1$ and $h$. Coefficients $\delta_1$ and $\delta_2$ will be chosen by the analysis similar to the one we demonstrated in CDFFJS15-0.38. However, the exact form of $\delta_1$ and $\delta_2$ is complicated, so we omit it here.

An elaborate calculation shows that $(\hat{x},\hat{y})$ is an $\epsilon$-NE with $\epsilon=(1/2-\sqrt{6}/18)\approx 0.364$.

\subsubsection*{DMP07-0.38}

The main idea of DMP07-0.38 is to find two strategy profiles $(\alpha,\beta)$ and $(x,y)$ approximating the values of a Nash equilibrium and then make a proper mixture of these strategy profiles. To guess the value pair of a Nash equilibrium, the first step is to discretize the value set (i.e., interval $[0,1]$) into set $\mathcal{V}=\{0,\epsilon,\dots,k\epsilon\}$ with $k\epsilon\leq 1<(k+1)\epsilon$, where $\epsilon>0$ is the input tolerance. Then try each pair of values $v_R,v_C\in\mathcal{V}$ of a Nash equilibrium to the two players. Note that there are in total $O(1/\epsilon^2)$ possible pairs. For each guess, do the search-and-adjust procedure. The search phase is as follows.
\begin{enumerate}
    \item Approximately secure a pairs of value  $(v_R,v_C)$ by $k$-uniform strategy profiles $(\alpha,\beta)$ with $k\geq\epsilon^2/4$ so that $\alpha^\T R\beta\geq v_R-3\epsilon/2$ and $\alpha^\T C\beta\geq v_C-3\epsilon/2$.
    \item Find a strategy profile $(x,y)$ approximating $v_R$ and $v_C$ as the feasible solution on $\Delta_m\times\Delta_n$ of the following linear program:
        \begin{align*}
        \alpha^\T Ry&\geq v_R-3\epsilon/2,\\
        R^iy&\leq v_R+\epsilon/2, i=1,\dots,m,\\
        x^\T R\beta&\geq v_R-3\epsilon/2,\\
        \alpha^\T Cy&\geq v_C-3\epsilon/2,\\
        x^\T C_j&\leq v_C+\epsilon/2, j=1,\dots,n,\\
        x^\T C\beta&\geq v_C-3\epsilon/2.
        \end{align*}
\end{enumerate}
Now we have got strategy profiles $(\alpha,\beta)$ and $(x,y)$. 

In the mixing phase, the algorithm constructs a strategy profile $(\hat{x},\hat{y})=(\delta\alpha+(1-\delta) x+\delta\beta+(1-\delta)y)$ with $\delta=3/2-1/\max\{v_R,v_C\}$. Then it outputs the one with the minimum approximation between $(\hat{x},\hat{y})$ and $(x,y)$. Similar to the situation in CDFFJS15-0.38, $\delta$ is chosen by the analysis process. Such a choice guarantees that the algorithm outputs a $(\gamma+\epsilon)$-NE with $\gamma=(3-\sqrt{5})/2\approx 0.382$.

As a remark, the algorithm runs for time $\max\{n,m\}^{O(\epsilon^{-2})}$ to get a $(0.382+\epsilon)$-NE, which is polynomial time in $m$ and $n$. On the other hand, other algorithms above clearly run in polynomial-time of $n$ and $m$, without auxiliary $\epsilon$.

\subsubsection*{TS methods (TS07-0.3393 and DFM22-1/3)}

Unlike the previous three algorithms, the TS methods apply an optimization approach in the search phase. Define the objective function as
\[f(x,y):=\max\{f_R(x,y),f_C(x,y)\},\quad (x,y)\in\Delta_m\times\Delta_n,\]
where $f_R(x,y)=\max(Ry)-x^\T Ry$ and $f_C(x,y)=\max(C^\T x)-x^\T Cy$.

It is immediate that the optimization procedure directly minimizes the approximation. However, function $f$ is nonconvex and nonsmooth. Thus the TS methods only try to find a local minimum point, or more precisely, a \emph{stationary point}. For differentiable functions, a stationary point is usually defined as a point with a zero gradient. However, function $f$ has non-differentiable points at which the gradient does not even exist. Nevertheless, function $f$ has directional derivatives at all points in all directions. Thus, we can alternatively define $(x,y)$ as a stationary point if all directional derivatives of $f$ at $(x,y)$ are nonnegative. Formally, it can be stated as
\[Df(x,y,x',y')\geq 0\text{ for all }(x',y')\in\Delta_m\times\Delta_n,\]
where $Df(x,y,x',y')$ is the Dini directional derivative defined as the following limit:
\[Df(x,y,x',y'):=\lim_{\alpha\downarrow 0}\frac{f(x+\alpha(x'-x),y+\alpha(y'-y))-f(x,y)}{\alpha}.\]

To find a stationary point, the TS methods use a descent procedure. For each round of optimization, follow the procedure below.
\begin{enumerate}
    \item Equalize $f_R$ and $f_C$ by changing $x$ or $y$ solely. It can be accomplished by a linear program.
    \item Find $(x',y')$ minimizing the Dini derivative $Df(x,y,x',y')$. It can also be completed by a linear program. 
    \item If $Df(x,y,x',y')<0$, do a line search and find a proper step size $\epsilon^*$. Move to the new point $(x+\epsilon^*(x'-x), y+\epsilon^*(y'-y))$.
\end{enumerate}

If $Df(x,y,x',y') \geq 0$, the procedure finds a stationary point and thus terminates. However, we cannot guarantee the algorithm terminates after a finite number of rounds. So we require the algorithm to terminate after finding a $\delta$-stationary point $(x,y)$, i.e., $Df(x,y,x',y') \geq -\delta$. It is proved that the descent procedure either finds a $\delta$-stationary point in $O(\delta^{-2})$ rounds or finds an infinite point sequence $(x_k,y_k)$ with $f(x_k,y_k) \to 0$ in an exponential speed.\footnote{It is claimed in the original paper \cite{tsaknakis2008optimization} that the descent procedure finds a $\delta$-stationary point in $O(\delta^{-2})$ rounds. However, it can be checked that what they actually proved is the statement here. In view of approximations, it is a better result than what they claimed.}

Apart from finding a ($\delta$-)stationary point $(x^*,y^*)$, there is a byproduct of the search phase. The linear program in step (2) has a dual linear program. The dual optimal solution is $(\rho^* w^*,(1-\rho^*)z^*)$, where $\rho^* \in [0,1]$, $w^*$ is a best response against $y^*$ and $z^*$ is a best response against $x^*$. Strategy pair $(w^*,z^*)$ is the key component in the mixing phase.

Now we turn to the mixing phase. For TS06-0.3393, it constructs strategy profiles $(\alpha x^*+(1-\alpha)w^*,z^*)$ and $(w^*,\beta y^*+(1-\beta)z^*)$ with proper coefficients $\alpha$ and $\beta$. Then it outputs the one with the minimum approximation among $(x^*,y^*)$ and these two constructed strategy profiles. Using inequalities generated by the linear programming in step (2) of the search phase, it can be shown that TS06-0.3393 finds a $(b+\delta)$-NE in time $O(\text{poly}(m,n,\delta^{-1}))$, where $b \approx 0.3393$ is the smallest real solution of the equation $4b(1-b)(1+b^2)=1$.

Evidence from empirical studies \cite{DBLP:conf/wea/FearnleyIS15,tsaknakis2008performance} suggests that TS06-0.3393 reaches an approximation far better than $0.3393$. It was conjectured that the upper-bound analysis of the approximation is not tight. However, \cite{DBLP:conf/sagt/ChenDHLL21} in 2021 proved that there exists a game instance such that TS06-0.3393 is able to reach its tight bound. Even worse, no matter how to choose $\alpha$ and $\beta$, bound $b \approx 0.3393$ will be attained by the strategy profile $(\alpha x^*+(1-\alpha)w^*, \beta y^*+(1-\beta)z^*)$ on that game. A stronger version of this tightness was proved by \cite{smoothed-tight-TS} in 2022. They showed that there exists a game instance such that when the descent procedure starts from a random initial point, the bound $b$ is reachable with a positive probability by all choices of $\alpha$ and $\beta$.

Such strong negative results indicate that we need to jump out of the ``square'' $(\alpha x^*+(1-\alpha)w^*, \beta y^*+(1-\beta)z^*)$ and make mixings in a larger space. That leads to DFM22-$1/3$. Specifically, the mixing is divided into five cases. In two of the cases, the algorithm outputs $(x^*,y^*)$. In one of the cases, the algorithm outputs $(w^*,z^*)$. For the two remaining cases, since they are symmetric, we here only introduce one. Define a new strategy $\hat{y}=(z^*+y^*)/2$ and take a best response $\hat{w}$ against $\hat{z}$. Construct strategy profiles $(p w^*+(1-p)\hat{w}^*,z^*)$ and $(w^*,(1-q)\hat{y}+qz^*)$ with $p$ and $q$ properly chosen from the algorithm analysis. Then the algorithm outputs the strategy profile with the minimum approximation among $(x^*,y^*)$ and these two constructed strategies. 

By a complicated and technical analysis, it was shown that DFM22-$1/3$ can find a $(1/3+\delta)$-NE in time $O(\poly(m,n,\delta^{-1}))$.

\begin{remark}\label{remark:extend-TS}
Indeed, the generality of optimization approaches allows TS methods to be applied to calculate NE in other kinds of games. In 2011, Kontogiannis and Spirakis \cite{DBLP:conf/wea/KontogiannisS11} proposed a polynomial-time algorithm computing $(1/3+\delta)$-approximate symmetric Nash equilibria in symmetric games by quadratic programming. It was proved that a stationary point is a $1/2$-NE \cite{DBLP:conf/sagt/ChenDHLL21,tsaknakis2008optimization}. Thus, if we do not implement the mixing step, we can find a $(1/2+\delta)$-NE. Such an idea was generalized by \cite{DBLP:journals/algorithmica/DeligkasFSS17} in 2017. They presented a polynomial-time algorithm calculating a $(1/2+\delta)$-NE in polymatrix games, which is the only known algorithm in the literature computing nontrivial NE of polymatrix games. Also, \cite{tsaknakis2008optimization} proved that in zero-sum bimatrix games, all stationary points are Nash equilibria. Thus TS06-0.3393 will find a $\delta$-NE. In win-lose bimatrix games (and games such that $R_{ij}+C_{ij}\leq 1$ for all $i,j$), any stationary point is a $1/4$-NE \cite{tsaknakis2008optimization}; hence TS06-0.3393 will find a $(1/4+\delta)$-NE.
\end{remark}

In summary, the search-and-mix methods are a powerful methodology to derive a polynomial-time algorithm finding NE with a constant approximation. However, the analysis of such methods becomes more complicated and cumbersome, while the approximation improvement seems little. New techniques would be required to break through the contemporary bottleneck.

\subsection{Finding \texorpdfstring{$\epsilon$}{eps}-WSNE: zero-sum-game-based methods}\label{subsec:epsWSNE}

At the first glance, WSNE is a similar notion to NE. However, they are essentially different (see \Cref{ex:WSNE-diff}). Here we make a further examination. Consider a strategy profile $(x,y)$ and another strategy $w$ of the row player. We shift $x$ to $w$ a little and the result is $((1-p)x+pw,y)$, where $p\ll 1$. If $(x,y)$ is an $\epsilon$-NE, then the resulted strategy profile is an $(\epsilon+o(1))$-NE as $p\downarrow 0$. However, when $(x,y)$ is an $\epsilon$-WSNE, no matter how small $p$ is, a bad $w$ will make the resulted strategy profile have a well-supported approximation close to $1$. Hence mixing methods using convex combinations for NE usually do not work for WSNE.

The first attempt to calculate $\epsilon$-WSNE was given by Daskalakis, Mehta, and Papadimitriou \cite{daskalakis2009note} in 2006. By rounding the payoff up to $1$ or down to $0$, they show a natural reduction from a general bimatrix game to a win-lose game. An approximation scheme is then given based on such a reduction. Such an algorithm is proved to find a $5/6$-WSNE under some graph-theoretic conjecture. The first polynomial-time algorithm that is unconditionally proved to have a constant well-supported approximation was given by Kontogiannis and Spirakis \cite{kontogiannis2007efficient} in 2008. Their algorithm can find a $2/3$-WSNE. Improving their work, in 2015, \cite{DBLP:conf/wea/FearnleyIS15} presented an algorithm finding a $0.6607$-WSNE. In 2015, a novel distributed approach was discovered by \cite{DBLP:journals/algorithmica/CzumajDFFJS19}, which calculates a $0.6528$-WSNE. With the growing intricacy of these polynomial-time algorithms, the improvement is, however, incremental. In 2019, Fasoulakis and Markakis \cite{DBLP:conf/aaai/FasoulakisM19} showed that one could calculate a $(1/2+\delta)$-WSNE in time $n^{\log\log(n^{1/\epsilon})/\epsilon^2}$, indicating the possibility of further improvements on the running time to calculate a $(1/2+\delta)$-WSNE. Indeed, seven years later after \cite{DBLP:journals/algorithmica/CzumajDFFJS19}, Deligkas, Fasoulakis, and Markakis \cite{1/2-WSNE} gave an intuitive and simple polynomial-time algorithm computing a $(1/2+\delta)$-WSNE in 2022. Although these methods appear different, we can summarize them into zero-sum-game-based methods: Solve a zero-sum game, and if the well-supported approximation is unsatisfying, then adjust the probability distribution or find a certain sub-game (possibly a pure strategy profile).

Similarly, we name the algorithm in the form of author initials plus the publishing year and the well-supported approximation their algorithm guarantees. A summary of these polynomial-time algorithms is presented in \Cref{tab:epsWSNE}.
 
\begin{table}[ht]
    \centering
    \caption{Summary of polynomial-time algorithms computing $\epsilon$-WSNE with constant $\epsilon$. Many works have a journal version. Our citations here adopt the journal version, if possible. The approximation $x+\delta$ means that given $\delta>0$, one can calculate a $(x+\delta)$-WSNE in polynomial time of $m$ and $n$.}
    \label{tab:epsWSNE}
    \begin{tabular}{ccc}
    \toprule
         Author initials & Publish year & Well-supported approximation \\\midrule\midrule
         KS \cite{kontogiannis2007efficient} & 2007 & $2/3$ \\
         FGSS \cite{DBLP:journals/algorithmica/FearnleyGSS16} & 2012 & $0.6607$ \\
         CDFFJS \cite{DBLP:journals/algorithmica/CzumajDFFJS19} & 2015 & $0.6528$ \\
         CFM \cite{DBLP:conf/sagt/CzumajFJ14} & 2014 & $1/2+\delta$ (symmetric games only) \\
         DFM \cite{1/2-WSNE} & 2022 & $1/2+\delta$ \\\bottomrule
    \end{tabular}
\end{table}

\subsubsection*{KS07-2/3}

The basic idea is to convert the original game $(R,C)$ into a proper zero-sum game lying between $(R,-R)$ and $(C,-C)$. Let $Z=-(R+C)$. To find such a game, a parameterized method is applied: Consider the zero-sum game $(R+\delta Z,-(R+\delta Z))$ with the parameter $\delta\in[0,1]$. Suppose $(\bar{x},\bar{y})$ is a Nash equilibrium of the game $(R+\delta Z,-(R+\delta Z))$ (which can be found in polynomial time). Then direct calculations show that $(\bar{x},\bar{y})$ is a $\epsilon(\delta)$-WSNE with $\epsilon(\delta)=\max\{\delta,1-\delta\}(\max_{i,j}Z_{ij}-\min_{i,j} Z_{ij})$. An optimal choice of $\delta$ is $1/2$.

Note that $\epsilon(1/2)$ could be equal to $1$, so only using $(\bar{x},\bar{y})$ does not work. However, we can control the approximation by setting an upper bound for payoff values. If the maximum payoff value of all pure strategy profiles is no smaller than $1-\zeta$ for both players, then there exists a pure strategy profile being a $\zeta$-WSNE. So we consider the remaining case that either $R_{ij}<1-\zeta$ for all $i,j$ or $C_{ij}<1-\zeta$ for all $i,j$. Then $\epsilon(1/2)$ is controlled by $1/2\cdot(2-\zeta)$. Taking both cases into consideration, we get an upper bound of $\min\{\zeta,1-\zeta/2\}\leq 2/3$. When $\zeta=2/3$, the equality holds.

As a remark, in their original paper, this proper zero-sum game method is introduced by considering only win-lose games. There, by taking $\delta=1/2$, $(\bar{x},\bar{y})$ is a $1/2$-WSNE.

\begin{remark}
    Similar to the situations in $\epsilon$-NE, when we only consider supports of size $O(\sqrt[3]{\log n})$, $2/3$ is the minimum possible well-supported approximation one can reach \cite{anbalaganPolylogarithmicSupportsAre2013}.
\end{remark}

\subsubsection*{CFM14-1/2 (symmetric games only)} 

Although CFM14-1/2 does not follow the basic idea of the zero-sum-game-based method, the novel concepts in CFM14-1/2 are important parts of the state-of-the-art algorithm DFM22-1/2, so we present this algorithm here. These key concepts are \emph{preventing exceeding payoffs} and \emph{well-supporting payoffs}. We begin by introducing these concepts.

A strategy $x$ of the row player is said to \emph{prevent exceeding payoff} $u$ if the column player payoff of the best response to $x$ does not exceed $u$. A similar definition applies to the column player. A strategy profile $(x,y)$ is said to prevent exceeding payoff $(v,u)$ if $x$ prevents exceeding $u$ and $y$ prevents exceeding $v$. Note that when $(v,u)$ is given, we can compute a strategy profile $(x,y)$ preventing exceeding payoff $(v,u)$ by linear programming. There is a direct connection between preventing exceeding payoffs and WSNE: If $(x,y)$ prevents exceeding payoff $(v,u)$, then $(x,y)$ is a $\max\{v,u\}$-WSNE.

To give another upper bound on the well-supported approximation, we need the notion of well-supporting payoffs. On the contrary to preventing exceeding payoffs, well-supporting payoffs characterize the payoff a player can secure on any pure strategy in the support of their own strategy. Formally, a strategy $x$ of the row player is said to \emph{well support payoff} $v$ against $y$ if $(e_m^i)^\T Ry\geq v$ holds for every $i\in\supp(x)$. A similar definition applies to the column player. A strategy profile $(x,y)$ is said to well support payoff $(v,u)$ if $x$ well supports $v$ and $y$ well supports $u$. 

An analog connection exists between well-supporting payoffs and WSNE: If $(x,y)$ prevents exceeding payoff $(v,u)$, then $(x,y)$ is a $(1-\min\{v,u\})$-WSNE. However, since the supports are unknown, it is not very direct how to use linear programming to find such a strategy profile. Fortunately, if $(v,u)$ is well supporting payoffs of some Nash equilibrium $(x^*,y^*)$, it is proved that we use a $\kappa(\delta)$-uniform strategy profile to find a strategy profile $(x,y)$ which well supports $(v-\delta,u-\delta)$, where $\kappa(\delta)=\lceil 2\delta^{-2}\log(\delta^{-1})\rceil$. An exhaustive search of such $\kappa(\delta)$-uniform strategy profiles will find a solution with approximately desired well-supporting payoffs.

Now, we combine these two notions to get CFM14-1/2. By Nash's fundamental work, there always exists a symmetric Nash equilibrium $(x^*,x^*)$. So the payoffs of both players are simultaneously no smaller than $1/2$ or greater than $1/2$. CFM14-1/2 follows the following procedure.
\begin{enumerate}
    \item Try to find a strategy profile that prevents exceeding the payoff $(1/2,1/2)$ by linear programming. If there is a feasible solution $(x_1,y_1)$, then $(x_1,y_1)$ is a $1/2$-WSNE.
    \item If there is no feasible solution, then there must exist a symmetric Nash equilibrium well that supports the payoff $(1/2,1/2)$. Given $\delta>0$, make an exhaustive search on $\kappa(\delta)$-uniform strategy profiles to find a pair $(x_2,y_2)$ which well supports the payoff $(1/2-\delta,1/2-\delta)$. Then $(x_2,y_2)$ is a $(1/2+\delta)$-WSNE.
\end{enumerate}

CFM14-1/2 clearly finds a $(1/2+\delta)$-WSNE. The running time of the algorithm is mainly determined by the exhaustive search, which takes $n^{O(\kappa(\delta))}$ trials with a polynomial running time in each trial. Thus the algorithm runs in polynomial time of $n$.

\subsubsection*{FGSS12-0.6607}

KS07-2/3 was designed based on how payoff values of $Z$ range. A finer discussion would produce a better approximation result. Indeed, FGSS12-0.6607 adopts two more observations and thus is able to calculate a $(2/3-\delta)$-WSNE with some $\delta>0$. To simplify the notations, we write $D=(R-C)/2$. Then the zero-sum game in KS07-2/3 is $(D,-D)$.

The first observation is shifting probabilities. Suppose $(x,y)$ is a Nash equilibrium of game $(D,-D)$. Suppose in the game $(R,C)$, the column player has a small regret on $(x,y)$ while the row player has a regret of $2/3$ on $(x,y)$. Note that if the column player changes their probability distribution over $\supp(y)$, their regret will not change. However, the row player could have less regret due to such changes. Thus shifting probabilities and concentrating on better strategies will decrease the well-supported approximation.

The second observation is matching-pennies sub-games. If any changes of the column player cannot affect the regret of the row player, then there must be a matching-pennies-like sub-game of $(D,-D)$ over the support of the column player. So simply playing the Nash equilibrium of this sub-game will produce a $1/2$-WSNE of game $(R,C)$.

Utilizing these observations, FGSS12-0.6607 can be described in three procedures.
\begin{enumerate}
    \item Find the best pure $\epsilon$-WSNE. This is a generalization of the first step of KS-2/3, setting no assumption on payoff values.
    \item Find the best pure $\epsilon$-WSNE with $2\times 2$ support. This step corresponds to the observation of matching-pennies sub-games. Note that when the support is given, optimizing the well-supported approximation becomes a linear program.
    \item Find the Nash equilibrium $(x,y)$ of the zero-sum game $(D,-D)$ and find the best $\epsilon$-WSNE supported on $\supp(x)\times\supp(y)$. This step follows the second step of KS07-2/3 and utilizes the observation of shifting probabilities. Both steps can be computed by linear programming.
\end{enumerate}
The algorithm then outputs the best $\epsilon$-WSNE among the found candidates in the three procedures.

By a parameterized method similar to KS07-2/3 and careful analysis of these two observations, it is proved that FGSS12-0.6607 yields a $(2/3-0.005913759)$-WSNE, or approximately a $0.6607$-WSNE. As for time complexity, there are a polynomial number of enumerations, and the time cost in each enumeration is polynomial (solving linear programming). Thus the algorithm takes a polynomial time.

\subsubsection*{CDFFJS15-0.6528}

In \Cref{subsec:epsNE}, it was mentioned that \cite{DBLP:journals/algorithmica/CzumajDFFJS19} gave a distributed method to compute NE. Their method, with proper refinements, can also be used to calculate $\epsilon$-WSNE. Their first step is to provide a base algorithm computing a $2/3$-WSNE via zero-sum games different from the one in KS07-2/3. Then, two observations similar to the ones in FGSS12-0.6607 are given. The improved algorithm is then proposed based on these two observations.

We first state the basic step of finding Nash equilibria in zero-sum games. Compute a Nash equilibrium $(x^*,y^*)$ for zero-sum games $(R,-R)$ and a Nash equilibrium $(\hat{x},\hat{y})$ for zero-sum games $(-C,C)$ respectively. The value secured by $x^*$ in $(R,-R)$ is $v_r$. Similarly, the value secured by $\hat{y}$ in $(-C,C)$ is $v_c$. Suppose without loss of generality that $v_c\leq v_r$.

The base algorithm can be divided into three procedures for three different cases of payoff values. The case division is accomplished by a parameterized method similar to the one used in \Cref{subsec:epsNE} and KS07-2/3. The base algorithm tries strategy profiles $(\hat{x},y^*)$ and $(x^*,y^*)$ respectively in two different cases. If both pairs are not $2/3$-WSNE, then there must be a pure strategy profile being a $2/3$-WSNE.

To modify the base algorithm, we need two observations: shifting probabilities and match-pennies sub-games. These observations quite resemble the ones in FGSS12-0.6607, with slightly different statements. Below we present the description of the modified algorithm. Here, $z\in(0,1)$ is some constant optimized by the analysis of the algorithm.

\begin{enumerate}
    \item Try $(\hat{x},y^*)$ and $(x^*,y^*)$ first. If either one produces a $(2/3-z)$-WSNE, output the desired one.
    \item To utilize shifting probabilities, find a best response column $j^*$ against $x^*$, then a proper strategy $x_B$ supported on $\supp(x^*)$ and produce strategy profile $(x_B,e_n^{j^*})$. Output $(x_B,e_n^{j^*})$ if it is a $(2/3-z)$-WSNE.
    \item If shifting probabilities do not work, then it turns to constructing a matching-pennies-like $2\times 2$ sub-game. Suppose $j'$ is a best response column against $x_B$. 
    \begin{enumerate}
        \item If there is some row $i\in\supp(x^*)$ making $(e_m^i,e_n^{j^*})$ or $(e_m^i,e_n^{j'})$ a $(2/3-z)$-WSNE, output such a strategy profile.
        \item Otherwise, find two rows $b$ and $s$ so that the sub-game formed by row $b,s$ and column $j^*,j'$ has payoff values properly approximating a matching pennies. Then find a proper strategy profile $(x_{mp},y_{mp})$ supported on this sub-game with guaranteed payoffs for both players. Output $(x_{mp},y_{mp})$.
    \end{enumerate}
\end{enumerate}

The analysis of this modified algorithm is fully parameterized by $z$. Thus, all occurrences of the word ``proper'' in the algorithm description should be determined by some expression of $z$ so that the well-supported approximation is $2/3-z$. The optimal choice of $z$ is a constant of a complicated expression, which is roughly equal to $0.013906376$. Hence the modified algorithm produces a $0.6528$-WSNE. By an argument similar to FGSS12-0.6607, it is clear that CDFFJS15-0.6528 runs in polynomial time.

\subsubsection*{DFM22-1/2}

DFM22-1/2 starts from zero-sum games the same as CDFFJS15-0.6528: Compute Nash equilibrium $(x^*,y^*)$ of game $(R,-R)$ and Nash equilibrium $(\hat{x},\hat{y})$ of game $(-C,C)$. Similarly, consider the strategy profile $(\hat{x},y^*)$. Now we discuss three different cases of payoff values.
\begin{enumerate}
    \item \emph{Low payoffs}. Note that when the column player plays $y^*$, i.e., minimizes the gain of the row player, the row player can secure at most $(x^*)^\T Ry^*$ payoff. So if the higher value secured by player $A$ when player $B$ minimizes the gain of player $A$ is low, no more gain can be obtained. Thus the well-supported approximation is low.
    \item \emph{Low-high payoffs}. If exactly one player, say $A$, benefits from deviating, then we let them concentrate on pure strategies which are played originally but yield a ``low'' maximum payoff for their opponent. The payoff of $A$ is high by deviating, while the opponent does not have more regrets after deviating. Note that this case resembles the observation of shifting probabilities in FGSS12-0.6607 and CDFFJS15-0.6528.
    \item \emph{High payoffs}. If both players have high payoffs, then there must be a sub-game so that any Nash equilibrium of this sub-game guarantees a payoff of at least $1/2$ for both players. To find a Nash equilibrium with such high payoffs in the sub-game, an exhaustive search of $k$-uniform strategies will be used. It is worth noting that a very similar technique is equipped in CFM14-1/2.
\end{enumerate}

Suppose $(x^*)^\T Ry^* \geq \hat{x}^\T C\hat{y}$. The above three cases correspond to the following three procedures:
\begin{enumerate}
    \item If $(x^*)^\T Ry^* \leq 1/2$, output $(\hat{x},y^*)$.
    \item If there exists a strategy $x'$ such that $\supp(x') \subseteq \supp(x^*)$, and $(x')^\T Ce_n^j \leq 1/2$ for every $j=1,\dots,n$, then return $(x',y^*)$.
    \item Otherwise, make an exhaustive search on $\kappa(\delta)$-uniform strategy profiles being $(1/2+\delta)$-WSNE, where $\kappa(\delta)=\lceil 2\delta^{-2}\log(\delta^{-1})\rceil$.
\end{enumerate}

These three cases guarantee DFM22-1/2 to find a $(1/2+\delta)$-WSNE. To compute the time complexity, note that the second procedure is a linear programming on $x'$ and the third procedure is a brute-force algorithm running in time $\max\{n,m\}^{O(\kappa(\delta))}$. So it runs in a polynomial time of $m$ and $n$.

Now we summarize all the methods computing an $\epsilon$-WSNE. They all start by solving zero-sum games related to the game $(R,C)$. Then they will take the following three tactics:
\begin{enumerate*}
    \item Pick a good pure strategy profile.
    \item Shift probabilities to concentrate on proper pure strategies.
    \item Find a sub-game and a proper strategy profile supported on this sub-game.
\end{enumerate*}
Compared with algorithms in \Cref{subsec:epsNE}, there are no more mixings using convex combinations. The mixing method is either to concentrate the probability distribution or simply take another strategy profile. Such a difference provides evidence that $\epsilon$-NE are essentially different from $\epsilon$-WSNE. Apart from the result of converting an $\epsilon^2/8$-NE to an $\epsilon$-WSNE, it almost remains blank how to build an underlying connection between these two approximation notions. On the other hand, parameterized methods are methodologies in common. Such methods play a key role in most of these approximation algorithms to get a (well-supported) approximation controlled by some constant.

\subsection{Heuristic algorithms: learning dynamics}\label{subsec:learning-dynamics}

We next come to another type of Nash equilibrium computation algorithms that are based on the heuristic idea of learning from repeated plays of the game. Players will make adaptive adjustments to their strategy through (infinitely) many rounds of the game. Theories of algorithms of this type mostly came after the initial design from heuristic ideas. These algorithms are implemented in an online form and adopted in certain machine learning methods.
All dynamic algorithms in this part are designed for multi-player normal-form games.

We first introduce some notations for multi-player games. Suppose there are $N$ players, denoted by $1,\dots, N$. Denote by $\Pi_i$ the set of all the pure strategies of player $i$ and let $n_i = |\Pi_i|$. The payoff function of player $i$ is $u_i:\Delta_{n_1}\times\dots\times\Delta_{n_N}\to\R$. We usually write it as $u_i(\bm{p}_i,\bm{p}_{-i})$ when player $i$ plays strategy $\bm{p}_i$ and other players play strategy profile $\bm{p}_{-i}$. The set of pure best response of player $i$ against $\bm{p}_{-i}$ is denoted by $\BR_i(\bm{p}_{-i})$. Now we are ready to present the algorithms.

\subsubsection*{Fictitious play (FP)}

Fictitious play (FP) is a classic learning dynamic method for Nash equilibrium computation. At each step of FP, players play a (pure) best response to the empirical distribution of their opponent’s past plays. The basic form of fictitious play was first given by Brown \cite{brown1951iterative} in 1951 though Cournot considered a quite similar best-response dynamic method in 1838 \cite{Cournot}. Ever since then, a number of variants have been devised. The most well-known ones are three variants of FP: discrete-time fictitious play (DTFP), continuous-time fictitious play (CTFP), and stochastic fictitious play (SFP). 
Detailed studies can be found in \cite{Nachbar2012}.
From an algorithmic perspective, DTFP is based on an iterative simulating method. We shall focus on DTFP in this paper. 

Let $a_i^t$ denote the pure strategy played by player $i$ in time $t$. The \emph{empirical distribution} of player $i$'s play \emph{before} time $t$, denoted by $p_i^t$, characterizes how frequently an action is taken in the past. It is defined as a probability measure over the set of pure strategies of player $i$. Specifically, the probability put on pure strategy $a_i$ is
\[p_i^t(a_i)=\frac{1}{t}\sum_{\tau=0}^{t-1}I(a_i^\tau=a_i),\]
where $I(a_i^\tau=a_i)$ is the indicator of $a_i^\tau=a_i$, which equals $1$ if $a_i^\tau=a_i$ and $0$ otherwise. Note that $p_i^t$ is also a mixed strategy of player $i$.

In DTFP, each player plays picks a pure strategy at time $0$. Alternatively, player $i$ may choose a mixed strategy $p_i^0$ as their beginning. At time $t$, player $i$ plays a pure best response against the product of the empirical distributions of all their opponents, i.e., pick $a_i^t\in\BR(\bm{p}_{-i}^t)$. 

Note that all players move simultaneously in each round. Thus, DTFP can be implemented in a distributed form. Also, it is naturally an online procedure. Although in DTFP, players put equal weights on players in each round, they can indeed choose weights differently, which produces different variants of DTFP. One example is the \emph{best-response dynamic}, in which all players put full weight on the play of the last round. In addition to choosing a pure strategy in each round, players can pick a mixed strategy based on the empirical distribution of their opponents. A well-studied variant is called smooth fictitious play (SFP). In SFP, we are given a nonnegative parameter family $\{\eta_t\}_{t=1}^\infty$, each player $i$ chooses a strategy $x^t$ by playing each strategy $j$ with probability proportional to $\exp\left(\eta_tu_i(e^j,\bm{p}_{-i}^t)\right)$.

Now we turn to the theory of DTFP. The convergence of DTFP is defined as the weak convergence of empirical distributions of all players. If all players do not alter their pure strategy after some rounds, then DTFP converges. Furthermore, a simple argument shows that DTFP must converge to a Nash equilibrium. With a bit more effort, one can show that if DTFP converges, then the limit distribution must be a Nash equilibrium. Thus for DTFP, the key problem of Nash equilibrium computing is: When does DTFP converge?

Brown conjectured that DTFP converges on two-player zero-sum games. The first systematic study on the analysis of DTFP was given by Robinson \cite{Robinson51} in 1951, which gave an affirmative answer to Brown's conjecture. Apart from two-player zero-sum games, Nachbar \cite{Nachbar90Evolutionary} showed that fictitious play converges when the game is solvable by iterated strict dominance in 1990. In the following, it has been shown that DTFP converges on $2\times n$ bi-matrix games by Metrick and Polak \cite{FPConvergence} in 1994. Also, Monderer and Shapley \cite{monderer1996fictitious} showed the convergence result on potential games in 1996. In 2022, \cite{chen2022convergence} provides a new sufficient condition on the convergence of DTFP: DTFP converges on any linear combination of a harmonic game (competition) and a potential game (cooperation) if they sum to be strategically equivalent to a zero-sum game or an identical interest game.

Computation does not allow infinite rounds of playing. Thus, we must stop before some finite rounds. If DTFP converges (to a Nash equilibrium), then, by definition, termination after a sufficiently large number of rounds will lead to an $\epsilon$-NE. So another main concern on DTFP is the convergence rate: How many rounds do we need to reach an $\epsilon$-NE? It was implied from Robinson's work that for two-player zero-sum games, one must find an $\epsilon$-NE after $\epsilon^{-\Omega(\max\{m,n\})}$ rounds. It was conjectured by Karlin~\cite{Karlin1960MathematicalMA} in 1959 that it will take $\epsilon^{-2}f^2(R)$ rounds to reach an $\epsilon$-NE, where $f(R)$ is some description complexity of the payoff matrix. However, Daskalakis and Pan \cite{DBLP:conf/focs/DaskalakisP14} in 2014 showed a counter-example to this conjecture. Specifically, they show that when the row player has an $n\times n$ identity matrix, there exists a dynamic in the game with a convergence rate as slow as $\Omega(t^{-1/n})$ when doing $t$ rounds of DTFP. For the lower bound, Brandt, Fischer, and Harrenstein \cite{DBLP:journals/mst/BrandtFH13} showed in 2010 that convergence could indeed take exponential rounds in symmetric two-player zero-sum games, even when the game is solvable via iterated strict dominance. They also showed an exponential convergence speed in most known classes of games on which DTFP converges, including non-degenerate $2\times n$ bimatrix games and symmetric zero-sum win-lose-tie games. Finally, for general bimatrix games, it was proved by Conitzer \cite{conitzer2009approximation} in 2009 that running $r$ rounds of DTFP yields a $(r+1)/(2r)$-NE, which is a tight bound. We mentioned in \Cref{subsec:epsNE} that $1/2$ is the best approximation one can reach by a constant number of supports. Thus, such an approximation result on DTFP is asymptotically optimal.

As for SFP, if we choose some proper parameters, the convergence rate in two-player zero-sum games is quite fast. Two independent works, \cite{Fudenberg1995ConsistencyAC} by economists Fudenberg and Levine in 1995 and \cite{Freund1999AdaptiveGP} by computer scientists Freund and Schapire in 1999, showed that if $\eta_t=\Theta\left(\sqrt{t}\right)$, then after $O(\log(m+n)\epsilon^{-2})$ rounds, the empirical distribution profile is an $\epsilon$-approximate Nash equilibrium.

\subsubsection*{No-regret learning dynamics.}

No-regret learning is a commonly used approach in algorithmic game theory for finding Nash equilibria in games. It operates as an online learning algorithm, where each player continually updates their strategy based on their past experiences and the experiences of others, without committing to a fixed strategy beforehand. The strategies are updated in a manner that minimizes the player's cumulative regret, which is calculated as the difference between the actual payoffs received and the payoffs that could have been obtained from playing an alternate strategy. Formally, the cumulative regret of player $i$ for a pure strategy $a_i \in \Pi$ at iteration $T$ is defined as $R^T_i(a_i) \coloneqq \sum_{t=1}^T u_i(a_i, \bm{p}^t_{-i}) - \sum_{t=1}^T u_i(a_i, \bm{p}^t_{-i})$. Some widely-used algorithms in the no-regret framework include regret matching~\citep{hart2000simple}, Hedge~\citep{auer1995gambling}, and multiplicative weight update~\citep{arora2012multiplicative}.

The regret matching algorithm keeps the cumulative regret and obtains the new strategy by normalizing all the positive regret of each pure strategy:
\begin{equation*}
    p_i^{T+1}(a_i) = 
    \begin{cases}
    \frac{\max\{0, R^{T}_i(a_i)\}}{\sum_{a_i' \in \Pi_i} \max\{0, R^{T}_i(a_i')\}}, & \text{if the denominator is positive}, \\
    |\Pi_i|^{-1}, & \text{otherwise}.
    \end{cases}
\end{equation*}

The Hedge algorithm keeps the cumulative payoffs of each pure strategy and obtains the new strategy by a softmax function:
\begin{equation*}
    p_i^{T+1}(a_i) = \frac{\exp\left(\sum_{t=1}^T u_i(a_i, \bm{p}^t_{-i})/\tau\right)}{\sum_{a_i'\in \Pi_i} \exp\left(\sum_{t=1}^T u_i(a_i', \bm{p}^t_{-i})/\tau\right)},
\end{equation*}
where $\tau > 0$ is the temperature parameter of softmax.

Similarly, multiplicative weight update maintains the weight of each pure strategy 
\begin{equation*}
    w^{T+1}_i(a_i) = w_i^T(a_i)\cdot(1-\epsilon)^{u_i(a_i, \bm{p}_{-i}^T)}.
\end{equation*}
Without loss of generality, we assume $u_i \in [0, 1]$.
Each player plays the strategy according to the distribution 
\begin{equation*}
    p_i^T(a_i) = \frac{w_i^T(a_i)}{\sum_{a_i' \in \Pi_i}w_i^T(a_i')}.
\end{equation*}

No-regret learning algorithms are proved to converge to (approximate) coarse correlated equilibrium in general-sum normal-form games~\citep{cesa2006prediction}.

In zero-sum bimatrix games, if both players have a cumulative regret of $\epsilon(T)$, the product of their empirical distribution $(\sum_{t=1}^T p_1/T, \sum_{t=1}^T p_2/T)$ is shown to converge to an $O(\epsilon(T)/T)$-NE.
For general-sum multi-player games, if each player has a cumulative regret of $\epsilon(T)$, it is known that the joint strategy empirical distribution converges to a coarse correlated equilibrium of the game, with a rate of $O(\epsilon(T)/T)$~\citep{daskalakis2021near, cesa2006prediction}.

\subsubsection*{Double oracle (DO) and policy space response oracles (PSRO)}

Double Oracle (DO)~\citep{mcmahan2003planning} algorithm aims to find a Nash equilibrium in large bimatrix games. 
The algorithm works by applying two separate oracles, one for each player. 
The first oracle provides a best response to the current strategy of the other player, and the second oracle provides a best response to the output of the first oracle. 
This process continues iteratively until a Nash equilibrium is reached, where neither player can improve their strategy by unilaterally changing it.

Formally, at each iteration $t$, DO keeps the current available population of pure strategies $\Pi^t = (\Pi^t_1, \dots, \Pi^t_N)$ where $\Pi^t_i \subseteq \Pi_i$, and solves the Nash equilibrium $\bm{p}^t$ of the subgame restricted to the strategies in $\Pi^t$. Afterwards, each player $i$ adds a best response $\BR_i(\bm{p}^t_{-i})$ to $\Pi_i^{t+1}=\Pi_i^t \cup \BR_i(\bm{p}^t_{-i})$ if $\Pi_i^t$ is not full.

The guarantee of convergence for DO is based on the existence of Nash equilibria in the game. 
While in the worst-case scenario, DO may require adding all pure strategies to the subgame, empirical results show that DO often terminates early in many games. 
This early termination is due to the ability of DO to quickly identify the ``most influential'' strategies and restrict the game to a smaller subgame, effectively reducing the computational complexity of the problem.

Policy Space Response Oracles (PSRO)~\citep{lanctot2017unified} is a generalized version of the Double Oracle algorithm that utilizes reinforcement learning to approximate best responses. In each iteration, PSRO maintains a population of policies instead of just pure strategies. The Nash equilibrium of the restricted subgame is computed using an arbitrary Nash equilibrium solver, such as fictitious play or regret matching, based on the empirical payoff matrix $u^{\Pi^t}$. The matrix consists of the average utility of each player when they play their corresponding policies. The players then use reinforcement learning to find their approximate best response and expand their policy populations.
PSRO can be applied to more complex games, such as Markov games~\citep{lanctot2017unified} and extensive-form games~\citep{mcaleer2021xdo}, and has demonstrated state-of-the-art performance in games like Starcraft II (AlphaStar~\citep{vinyals2019grandmaster}) and Stratego (Pipeline PSRO~\citep{mcaleer2020pipeline}).

\subsection{Other algorithms for multi-player games with theoretical guarantees}

Finally, we present various algorithms which are suitable for multi-player games. Actually, apart from brute-force search in \Cref{subsec:brute-force} and learning dynamics in \Cref{subsec:learning-dynamics}, there are very few algorithms, to our knowledge.

For exact NE, there is no known algorithm, to our knowledge, to compute it for general $r$-player games with $r\geq 3$.

For approximate NE, the Lemke-Howson algorithm can be extended for multi-player games \cite{wilsonComputingEquilibriaNPerson1971}. A heuristic improvement was proposed by \cite{blumContinuationMethodNash2006}. Similarly, many two-player algorithms can be extended to multi-player cases. Based on the invariance principle (see \Cref{subsec:brute-force}), an enumeration-based algorithm was devised by \cite{porterSimpleSearchMethods2008}. Also, if we reform NE computation into LCP (see \Cref{subsec:LH-LCP}) or fixed-point computation, then we can use the corresponding solvers to compute approximate NE, e.g., \cite{doupNewSimplicialVariable1987}. Not surprisingly, these algorithms can take exponential time, or even have no time-complexity guarantee.

If we consider special kinds of games, then faster algorithms are possible. If the number of actions is polynomial to the number of players, we can compute $\epsilon$-NE in time $N^{\log\log N}$, where $N$ is the input size \cite{babichenkoSimpleApproximateEquilibria2014}. The only known polynomial-time example is \cite{DBLP:journals/algorithmica/DeligkasFSS17}, which presents a polynomial-time algorithm computing $(1/2+\delta)$-NE in polymatrix games. See \Cref{remark:extend-TS} for further discussions. If we restrict the number of players, there is also a result that extends a $k$-player algorithm to a $(k+1)$-player algorithm: If we have an algorithm computing $\alpha$-NE for $k$-player games, then we have an algorithm computing $1/(2-\alpha)$-NE for $(k+1)$-player games \cite{bosse2010new}.

\section{Empirical comparisons}
\label{sec:experiment}

Usually, the study of an algorithm should consider two aspects: theoretical guarantees and practical performance. In this section, we focus on the empirical comparisons of different algorithms presented in \Cref{sec:algorithm}. Again, our main focus is NE approximations over bimatrix games. Since there are very few empirical studies in the literature to our knowledge, we first review these works.

\subsection{Empirical studies in the literature}

Empirical studies are roughly divided into two classes: the comparison of algorithms finding exact/arbitrary approximate NE \cite{codenottiExperimentalAnalysisLemkeHowson2008,gattiCombiningLocalSearch2012,blumContinuationMethodNash2006,porterSimpleSearchMethods2008} and the comparison of algorithms finding $\epsilon$-NE/$\epsilon$-WSNE with constant $\epsilon$ \cite{DBLP:conf/wea/FearnleyIS15,tsaknakis2008performance,DBLP:conf/wea/KontogiannisS11,chen_tightness_2023}. 

Works in the first class usually design new algorithms \cite{porterSimpleSearchMethods2008} or propose variants on Lemke-Howson algorithm \cite{codenottiExperimentalAnalysisLemkeHowson2008,gattiCombiningLocalSearch2012,blumContinuationMethodNash2006}, and then compare them with the original Lemke-Howson algorithm. These novel methods significantly speed up the NE computation. Particularly, the enumeration-based algorithm by \cite{porterSimpleSearchMethods2008} has the lowest running time among these algorithms.

Works in the second class either focus on a specific algorithm \cite{tsaknakis2008performance,DBLP:conf/wea/KontogiannisS11,chen_tightness_2023} or focus on algorithms in \Cref{subsec:epsNE} and \Cref{subsec:epsWSNE} at that time \cite{DBLP:conf/wea/FearnleyIS15}. They showed that these algorithms often perform much better than their theoretical worst-case guarantees and very differently on different kinds of games. It is concluded that TS07-0.3393 outperforms other algorithms in terms of approximation. The running time of these algorithms is also evaluated, showing that although TS07-0.3393 has the best performance, if we only require weaker approximate equilibria, then other algorithms can find one faster.

To compare these algorithms, we need a test-case benchmark. The standard benchmark is GAMUT \cite{nudelman2004run}, which includes various game scenarios from economics and theoretical computer science, mostly two-player cases. \cite{chen_tightness_2023} provided all tight instances for TS07-0.3393, which also can be regarded as a benchmark.

\subsection{Research questions for comparisons}

Now we begin our empirical comparison by raising the following natural questions for practical use:
\begin{enumerate}[label=\textbf{RQ\arabic*:}]
    \item \textbf{Practical Approximation Performances.} What are the actual approximation and well-supported approximation performances of different algorithms on different games?
    \item \textbf{Random vs. Structured.} How does an algorithm with theoretical guarantees perform with respect to approximations on random games versus structured games in GAMUT?
    \item \textbf{Efficiency.} How efficiently do different algorithms run on different games?
    \item \textbf{Precision Error.} For the algorithm which invokes a numerical solver (e.g., LP solver), will it encounter precision errors (e.g., float overflow) during its computation procedure?
\end{enumerate}

To make a fair comparison, we compare algorithms within similar categories:
\begin{enumerate*}
\item Enumeration-based algorithms and Lemke-Howson algorithm,
\item Algorithms with specific approximation guarantees,
\item Learning Dynamics.
\end{enumerate*}
Not all categories are suitable for all the above RQs. We will discuss the availability of these RQs when presenting the results.

\subsection{Experiment setup}

We evaluate the performance of enumeration-based algorithms in \Cref{subsec:brute-force}, Lemke-Howson algorithms in \Cref{subsec:LH-LCP}, five learning dynamics in \Cref{subsec:learning-dynamics}, six approximate NE algorithms in \Cref{subsec:epsNE}, and four approximate WSNE algorithms on three different benchmark scenarios for two-player games: random zero-sum games of three different sizes, random general games of three different sizes, and nine different scenarios from the GAMUT of three different sizes each. For each scenario, we report the empirical average approximation and well-supported approximation. The approximation is calculated as defined, while well-supported approximation is calculated after ignoring supports with probabilities less than $10^{-10}$.

All algorithms are implemented using Python, due to the language's widespread use in the machine learning community (e.g., PyTorch, TensorFlow) and the availability of widely used scientific-computing libraries (e.g., NumPy, SciPy). The algorithms and testing environments are implemented by the authors, except for the Lemke-Howson algorithm, which is obtained from the open-sourced NashPy library \cite{knight2018nashpy}. All algorithms are tested on a Linux machine with $48$ core Intel(R) Xeon(R) CPU (E5-2650 v4@2.20GHz).\footnote{Since the main bottleneck of the computation is the speed of floating operations on CPUs (floating point operations per second, FLOPS), not the memory latency, we here do not list the configuration of memory.} 


\subsubsection{Algorithms}

We implement enumeration-based algorithms and use the implementations of Lemke-Howson from NashPy, a library for computing Nash equilibria in bimatrix games. We do not implement the Lemke-Howson algorithm since it is more probable that people will use NashPy rather than implement it by themselves.

For learning dynamics, we evaluate fictitious play, hedge, multiplicative weight update with exponential function (MWU-exp), multiplicative weight update with linear function (MWU-linear), and regret matching. We do not include DO or PSRO as baselines, as they require a subgame NE solver or external updating weights, respectively.

For approximate NE algorithms, we evaluate KPS06-0.75, DMP06-0.50, CDFFJS15-0.38, BBM07-0.36, TS07-0.3393, and DFM22-1/3. We exclude DMP07-0.38 as it cannot finish computing even in some $3 \times 3$ random bimatrix games within one day.

For approximate WSNE algorithms, we evaluate KS07-2/3, FGSS12-0.6607, CDFFJS15-0.6528, and DFM22-1/2. We do not include CFM14-$1/2$ as it only works for symmetric games and requires more than one day to compute on a $3\times 3$ game.

The implementation details can be found in \appref{sec:imple_details}.

\subsubsection{Test cases}

We evaluate the performance of our algorithms on both random zero-sum bimatrix games\footnote{It means that all entries in $R$ are independently and uniformly sampled from $[0,1]$, and then we set $C=-R$.} and random general bimatrix games\footnote{Similarly, it means that all entries in both $R$ and $C$ are independently and uniformly sampled from $[0,1]$.} with three different sizes: $10 \times 10$, $100 \times 100$, and $1000 \times 1000$. The games are generated with $40$ random seeds for the $10 \times 10$ size, $20$ random seeds for the $100 \times 100$ size, and $10$ random seeds for the $1000 \times 1000$ size. In addition, we also perform experiments on the GAMUT benchmark. 
The specifics of the GAMUT experiments are outlined in \appref{sec:experi}.

\subsection{Results and analysis}

Now we present the experimental results and their analysis in accordance with the classification we provide in \Cref{sec:algorithm}.
The detailed results are presented in \appref{sec:experi}.

\subsubsection{Enumeration-based algorithms and Lemke-Howson algorithm}

The only meaningful questions for these algorithms are RQ3 and RQ4. We found that the support-enumeration algorithm runs fast on $10 \times 10$ random games, but times out in all other cases. On the other hand, the Lemke-Howson algorithm struggles with degenerate games and cannot find a proper strategy profile, even in $10 \times 10$ random games. A common approach to handle degenerate games is to perturb the payoff matrices. We tried this method, but it only helped in a few cases. We found that this is due to float overflow, not degeneration. Using high-precision representation eases the overflow problem to some extent, but the issue still persists even when using 128-bit floating points.

We conclude that both algorithms are not practical, especially for large games. The support-enumeration algorithm runs too slowly in most cases (RQ3), while the Lemke-Howson algorithm (in the Nashpy implementation) faces precision difficulties in many cases (RQ4).

\subsubsection{Algorithms with specific approximation guarantees}\label{subsec:approx-solver}

We evaluate algorithms that aim to compute $\epsilon$-NE (class A) and $\epsilon$-WSNE (class B), where $\epsilon$ is constant. The algorithms in class A proceed by searching and mixing. The data in parentheses represent the result of the strategy profile before mixing.

\textbf{Results of class A.}

For RQ1, we observe that algorithms with a more complex design tend to exhibit better approximation. Also, it is worth noting that most algorithms have a much better approximation performance than the theoretical upper bound, matching the results in \cite{tsaknakis2008performance, DBLP:conf/wea/FearnleyIS15}. However, their performance in terms of well-supported approximations, particularly on random general games, is in general far worse than approximations. This is because the well-supported approximation is a stronger solution concept than the approximation. See \Cref{sec:definition} and \Cref{subsec:epsWSNE} for theoretical discussions.

For RQ2, surprisingly, they perform better on structured games in GAMUT than on random games. This can be explained by the original aim of these algorithms: They only guarantee the \emph{worst-case} performance is not too bad but make no guarantee on the average cases. 

For RQ3, the running time has the following relations: KPS06-0.75 $\approx$ DMP06-0.50 $<$ TS07-0.3393 $\approx$ DFM22-1/3 $\ll$ BBM07-0.36 $<$ CDFFJS15-0.38. Such relations can be understood as follows. For KPS06-0.75 and DMP06-0.50, they take linear time, thus being the fastest. TS07-0.3393 and DFM22-1/3 are almost the same, which are based on optimization and the main time is spent on the descent process, which computes several LPs. Even though every round of the descent procedure involves three LPs, TS07-0.3393 and DFM22-1/3 are efficient perhaps for three reasons:
\begin{enumerate}
    \item The line search performs well, reducing the number of iterations needed. A similar phenomenon was also observed by \cite{DBLP:conf/wea/FearnleyIS15}, who measured the running time of TS07-0.3393 with two different line search methods.
    \item The LP in each round has few constraints.
    \item Random initialization happens to be a good initialization with a reduced number of iterations. This speculation is based on an empirical observation in \cite{Duan2023IsNash}: Appropriate initialization can significantly improve the convergence speed of both TS07-0.3393 and DFM22-1/3. Thus, random initialization may help too.
\end{enumerate}
Note that the overall efficiency result is very different from the one in \cite{DBLP:conf/wea/FearnleyIS15}. The most possible reason is that the LP solver is very different in their implementations.

For RQ4, although all but KPS06-0.75 and DMP06-0.50 need to invoke the LP solver, only CDFFJS15-0.38 encountered precision error in a special scenario in GAMUT. This is because these algorithms solve different linear programs, and only very special constructions can cause an algorithm to fail.

Finally, we also study the effect of mixing phases. Only the CDFFJS15-0.38 algorithm had a performance that was worse on the output strategy profile than the strategy profile before mixing. This was due to its "mix-over-threshold" strategy, where mixing only occurs when the estimated approximation of the original strategy profile is high. However, this can lead to a worse strategy profile if the original strategy profile performs well in practice. Also, it is worth mentioning that experiments showed that CDFFJS15-0.38 (before mixing) and BBM07-0.36 find a Nash equilibrium in zero-sum games.\footnote{It is not hard to prove it theoretically. To focus on the main results, we omit it here.}

\textbf{Results of class B.}

For RQ1, CDFFJS15-0.6528 and DFM22-1/2 perform well in zero-sum games. This is because they have already output before the exhaustive search starts. Again, their practical well-supported approximations are far better than the theoretical guarantee. In every instance, the well-supported approximation of KS07-2/3 is always given by its output pure strategy profiles.

For RQ2, again, the performance on structured games is better than random games. We also attribute this to the nature of \emph{worst-case-guaranteed} design.

For RQ3, the running time has the following relations: KS07-2/3 $<$ CDFFJS15-0.6528 $\ll$ DFM-1/2 $\ll$ FGSS12-0.6607. However, these algorithms run much slower than those in class A; even worse, the latter two usually time out. FGSS12-0.6607 times out in games with large sizes due to an exhaustive search for finding the best pure $\epsilon$-WSNE with $2\times 2$ support using very large LP. DFM22-1/2 also times out for all general games when the third procedure, an exhaustive search on $k$-uniform strategy profiles, is entered, requiring very intractable computations. CDFFJS15-0.6528 solves two zero-sum games can only be completed with the interior method as the LP solver. The fastest algorithm in class B is KS07-2/3. Despite including a procedure for solving a zero-sum game, it mainly checks pure strategies due to its early stop strategy.

For RQ4, although all algorithms in class B invoke the LP solver, only CDFFJS15-0.6528 encountered precision errors. This is only because other algorithms solve LP of small sizes or simply time out.

\subsubsection{Learning dynamics}

For RQ1, unlike algorithms with theoretical guarantees, there is no unified description of the approximation performance. We observe that the average approximation becomes larger with increasing matrix size. In each scenario, the last iteration strategies are inferior to the average policy. The type of games appears to be the most significant factor impacting the approximation. For example, the average policy approximation in random general games is approximately ten times higher than in random zero-sum games. However, the last iteration strategies' approximation in random general games is relatively lower than in random zero-sum games. In random general games, the approximation for both average policy and last iteration strategies is comparable, but Hedge and regret matching algorithms outperform others significantly in random zero-sum games for last iteration strategies. Hedge algorithms are similar to MWU-exp, with the main difference being the updating rate -- Hedge uses $\log(2)/T$, while MWU-exp uses a general rate of $0.5$ (as does MWU-linear). When the updating rate of Hedge is set to $0.5$, its performance becomes comparable to that of MWU-exp. This highlights the significant impact of the updating rate on the performance of last iteration strategies in Hedge or MWUs for zero-sum games, but a lesser impact on average policy. On the other hand, regret matching does not require setting parameters, and its performance improves with increasing game size.

Then we turn to the results of the well-supported approximation, denoted by $\epsilon$. It should be noted that these learning dynamics have no guarantee on well-supported approximations. Again, we see the significant impact on the type of games. For random general games, the average policies have large $\epsilon$ values, but last-iteration strategies tend to perform better. The results of the algorithms are similar, except for Hedge. For $100 \times 100$ and $1000 \times 1000$ scenarios, Hedge's last iteration strategies underperform compared to others, but its average policies outperform. The performance of Hedge and MWU-exp can be improved by controlling the updating rate carefully. For the results on random zero-sum games, the pattern is similar to that of NE results. Hedge and regret matching still perform better for last-iteration strategies. However, for average policies, fictitious play surprisingly outperforms other algorithms on $10 \times 10$ and $100 \times 100$.

For RQ2, we again observed that learning dynamics perform better in most of these structured scenarios than random games. Indeed, they converge to the pure Nash equilibria in most cases. However, the hardest scenario appears to be the covariant game (CG) of size $1000$, where the dynamics struggle to reach a $0.001$-approximate Nash equilibrium, likely due to the absence of a pure Nash equilibrium in covariant games and the need for more steps to reach a mixed Nash equilibrium.

RQ3 does not have much meaning for the learning dynamics in our experiments since their running time highly relies on the number of iterations, which is set to be a constant. In addition, since there is no numerical solver at all, RQ4 is not applicable here too.
\section{Conclusion and discussion}
\label{sec:conclusion}

In this survey, we present two messages on Nash equilibrium computation, especially for bimatrix games.
\begin{enumerate}
    \item \textit{Theoretical aspect}. We classify most approximate Nash equilibrium algorithms with full theoretical guarantees into search-and-mix methods and zero-sum-game-based methods. It seems very hard to improve the (well-supported) approximation by refining such methods. New techniques are probably required.
    \item \textit{Empirical aspect}. Even if an approximate Nash equilibrium algorithm for bimatrix games does not have good enough theoretical guarantees, it could have far better performance in practice. Furthermore, although the series of algorithms in the literature has improving approximation bounds (including algorithms with a specific approximation guarantee and learning dynamics), they do not obtain the expected performances on random games.
\end{enumerate}

We also provide suggestions for the implementations and uses of algorithms in \Cref{subsec:epsNE}, \Cref{subsec:epsWSNE}, and \Cref{subsec:learning-dynamics}.
\begin{enumerate}
    \item For algorithms with a specific approximation guarantee (\Cref{subsec:epsNE} and \Cref{subsec:epsWSNE}), most of these approximation algorithms involve linear programming subroutines. Implementations of these algorithms should carefully handle numerical difficulties, especially those raised by the LP subroutines. Specifically, one should cautiously choose the LP solver making a trade-off between precision and running time. Evidence shows that by doing so, the emergence of numerical difficulty will almost vanish. 
    \item For learning dynamics (\Cref{subsec:learning-dynamics}), compared with results on random games, learning dynamics perform far better on structural games, including game scenarios in GAMUT (even covariant games) and games with (near) pure Nash equilibria. Thus, it is preferred to use learning dynamics on structural games. 
\end{enumerate}

Alternatively, when the game is sampled from a specific distribution, employing function approximation methods may prove advantageous~\citep{Duan2023IsNash, marris2022turbocharging, duan2023equivariant}. These methods involve the construction of a parameterized function as a Nash equilibrium approximator and its training through a standard machine learning pipeline. The Nash equilibrium approximator can subsequently expedite the computation of Nash equilibria. Notably, it has been demonstrated in \cite{Duan2023IsNash} that the outputs of Nash equilibrium approximator can also serve as effective initialization for TS06-0.3393 and DFM22-1/3, thus speed up the convergence of the two algorithms.

\subsubsection*{Open problems and future directions.} 
From a theoretical consideration, we raise the following open problems.
\begin{enumerate}
    \item Improve the $1/3$ bound for approximation via search-and-mix methods.
    \item Extend TS methods to multi-player games with a fixed number of players.
    \item Average-case approximation analysis of search-and-mix methods.
    \item Find new bimatrix-game classes whose Nash equilibria can be computed in polynomial time, just like zero-sum games.
    \item Determine new classes of games on which a dynamic (e.g., fictitious plays and no-regret learning dynamics) converges.
    \item Develop variants of learning dynamics with better convergence rates. 
    \item Develop polynomial-time approximation algorithms for general multi-player games.
    \item Develop an algorithm computing exact NE for multi-player games, or prove that it is beyond the computational power of Turing machines.
\end{enumerate}
   
On the other hand, possible research directions are also raised from an empirical perspective. 
\begin{enumerate}
    \item Overcome numerical difficulties caused by LP solvers and reduce large-size LP subroutines for algorithms with a specific approximation guarantee.
    \item Accelerate convergence of learning dynamics in a heuristic way.
    \item Design a machine-learning-based classifier that chooses the best algorithm when given a certain game.
    \item Propose a bimatrix-game benchmark that better distinguishes different algorithms.
    \item A complete algorithm library for NE computation. This can serve as the basic toolkit for practical use, especially for multi-agent reinforcement learning.
\end{enumerate}

\section*{CRediT authorship contribution statement}

\textbf{Hanyu Li:} Writing - Original Draft, Writing - Review \& Editing, Investigation, Software. \textbf{Wenhan Huang:} Writing - Original Draft, Writing - Review \& Editing, Investigation, Software. \textbf{Zhijian Duan:} Writing - Original Draft, Writing - Review \& Editing. \textbf{David Henry Mguni:} Writing - Review \& Editing, Project administration. \textbf{Kun Shao:} Supervision. \textbf{Jun Wang:} Conceptualization. \textbf{Xiaotie Deng:} Conceptualization, Writing - Review \& Editing, Funding acquisition.

\section*{Declaration of competing interest}

The authors declare that they have no known competing financial interests or personal relationships that could have appeared to influence the work reported in this paper.

\section*{Data availability}

No data was used for the research described in the article.

\section*{Acknowledgements}

This work was supported by Science and Technology Innovation 2030 - ``The New Generation of Artificial Intelligence'' Major Project (No. 2022ZD0114904).

\bibliographystyle{plain}
\bibliography{ref}

\appendix

\newpage
\section{Implementation Details}
\label{sec:imple_details}

\subsection{Algorithms}

The general implementation details for algorithms can be found in \Cref{tab:algo_general_details}.
We use $64$-bit floating number (\texttt{Numpy.float64}) for general calculation, and $128$-bit floating number (\texttt{Numpy.float128}) is only used for precision checking of a certain procedure, e.g., the linear programming, or the Lemke-Howson algorithm.

\begin{table}[ht]
    \centering
    \caption{Implementation details related to Python packages.}
    \label{tab:algo_general_details}
    \begin{tabular}{ll}
        \toprule
        Description & Specification\\
        \midrule
        \midrule
        Python version & Python 3.9\\
        NumPy version & 1.22.3\\
        SciPy version & 1.9.1\\
        LP solver & \texttt{scipy.linprog}\\
        NashPy version & 0.0.35\\
        Floating data type & \texttt{float64}\\
        \bottomrule
    \end{tabular}
\end{table}

Notice that several algorithms need to solve some linear programming (LP) as a sub-procedure. Due to the time limit and the precision problem, different algorithms need different LP solvers. We list the types of LP solvers from \texttt{scipy.linprog} used for each algorithm in \Cref{tab:algo_lp_type}.

\begin{table}[ht]
    \centering
    \caption{Implementation details related to LP solvers. \texttt{highs-ds} is an implementation of the dual revised simplex, and \texttt{highs-ipm} is an implementation of the interior-point method.}
    \label{tab:algo_lp_type}
    \begin{tabular}{ll}
        \toprule
        Algorithm & LP solver\\
        \midrule
        \midrule
        DFM22-$1/3$ & \texttt{highs-ds} \\
        FGSS12-$0.6607$ & \texttt{highs-ds} \\
        TS07-$0.3393$ & \texttt{highs-ds}\\
        BBM07-$0.36$ & \texttt{highs-ipm}\\
        CDFFJS15-$0.38$ & \texttt{highs-ipm}\\
        CDFFJS15-$0.6528$ & \texttt{highs-ipm}\\
        DFM22-$1/2$ & \texttt{highs-ipm}\\
        \bottomrule
    \end{tabular}
\end{table}

Then, we list the special setup for different algorithms.

\begin{itemize}[fullwidth]
    \item \emph{Enumeration-based algorithms and Lemke-Howson algorithm}. As discussed in \cite{DBLP:conf/wea/FearnleyIS15}, the approximation algorithm TS06-0.3393 has good performance in terms of both running time and approximation ratio. Hence, we compare the running time of the exact algorithms with that of TS06-0.3393. The experiments are conducted on both random general games and zero-sum games as specified in the setup.
    
    To ensure a fair comparison, we warm up the cache before testing an algorithm on a game. Specifically, each algorithm runs twice on the same game, and the running time of the second run is measured. To account for the exponential running time of exact algorithms, we set a timeout for each run. The timeout is set to be the maximum of $60$ seconds and the running time of TS06-$0.3393$ on the same game.

    \item \emph{Algorithms computing $\epsilon$-NE with constant $\epsilon$}. We measure the approximation and well-supported approximation of the output strategy profiles, as well as the strategy profiles before mixing. The methods KPS06-0.75 and DMP06-0.50 are exceptions, as their two-phase divisions are trivial. For precision requirements, the LP solver in TS07-0.3393 and DFM22-1/3 is the revised simplex algorithm. 

    \item \emph{Algorithms computing $\epsilon$-WSNE with constant $\epsilon$}. We optimize FGSS12-0.6607 using an early stop method. The method first tries pure strategy profiles and stops if a Nash equilibrium is found. The zero-sum game procedure and $2\times 2$ sub-game procedure are then tried.

    \item \emph{Learning dynamics}. We conduct $T=10^5$ iterations with zero initialization.
\end{itemize}

We use zero initialization rather than random initialization for algorithms (if they need) as their standard versions, e.g., some learning dynamics, except TS07-$0.3393$ and DFM22-$1/3$.
For TS07-$0.3393$ and DFM22-$1/3$, they need a non-zero strategy as the initialization strategy. Typically, there are two solutions, the average strategy or a random strategy.
As a fixed initialization strategy could meet the worst-case approximation~\citep{DBLP:conf/sagt/ChenDHLL21}, we choose to use a random one for them.

\subsection{Test cases}

A bimatrix game contains the following information.

\begin{itemize}
    \item Number of players: In this paper, we only consider $2$ players.
    \item Action space: The action space of a game is the joint action space of its players.
    \item Utility query: Player $i$ can only query its value given the joint action or its value vector given the joint action of other player(s).
\end{itemize}

We assume that a player cannot obtain the feedback of a strategy in a query, as in some realistic games.
For approximation algorithms that need to input the utility matrix, we assume the algorithms query each joint action in the whole action space once at the very beginning.

\subsection{Randomness}

The randomness in our experiments mainly comes from these three aspects:

\begin{itemize}
    \item \textit{Matrix game generator}. The random bimatrix games or GAMUT games are generated via some random seeds. To ease the influence of this randomness, we generate several bimatrix games for each game size and game type.
    \item \textit{Action choosing}. Some algorithms, e.g., regret matching, need to play a mixed strategy, but our environment requires one exact action. In this case, the action will be chosen randomly via its density. Some other algorithms might have several best-action candidates. In our experiments, the action will be chosen uniformly randomly among these candidates.
    \item \textit{Linear programming}. When solving a linear program is a sub-procedure of some algorithms, the linear program might have several solutions. We use a solution outputted by the \texttt{linprog} method from the \texttt{scipy} package, which we regard as a black-box function.
\end{itemize}
\newpage
\section{Experiment results}
\label{sec:experi}

In this appendix, we present the experimental results on both random games and GAMUT~\citep{nudelman2004run}.

In \Cref{tab:GAMUT}, we provide a brief description of the $9$ game distributions we used.
For each game distribution, we generate different instances by using different random seeds so that we randomly set the game parameters.
We achieve this by setting the `\texttt{-random params}' and `\texttt{-random seed}' flags.
For Covariant Game, the covariance $\rho$ is sampled uniformly from $[-1, 1]$.

\begin{table}[ht]
    \centering
    \caption{Descriptions of GAMUT classes.}
    \label{tab:GAMUT}
    \begin{tabular}{ll}
        \toprule
        Abbreviation & Full name\\
        \midrule
        \midrule
         BO & Bertrand Oligopoly \\
         CD & Cournot Duopoly \\ 
         CG & Covariant Game, $\rho \in [-1,1]$ \\
         GTD & Grab The Dollar \\
         GTTA & Guess Two Thirds Ave\\
         LG & Location Game \\
         MEG & Minimum Effort Game\\
         TD & Travelers Dilemma \\ 
         WOA & War Of Attrition\\
         \bottomrule 
    \end{tabular}
\end{table}

In the following sections, we present all experimental results in tables.

\subsection{Algorithms with specific approximation guarantees on random games}\label{subsec:experiment-approx-solver}

\begin{table}[H]
\centering
\caption{Approximation results for approximate NE solvers. We omit the publishing year in the algorithm name. Each datum in the table represents the average approximation of experiments on randomly generated games. The data in the parentheses is the average approximation before the mixing phase.}\label{tab:approxNE-of-approxNE-algo}
\begin{tabular}{cccccccc}
\toprule
\multicolumn{2}{c}{Scenario} & KPS-$0.75$ & DMP-$0.50$ & CDFFJS-$0.38$ & BBM-$0.36$ & TS-$0.3393$ & DFM-$1/3$\\
\midrule
\midrule
\multirow{6}{*}{Zero-Sum} & \multirow{2}{*}{$10$} & \multirow{2}{*}{0.4068} & \multirow{2}{*}{0.4280} & 0.2873 & \textbf{0.0000} & 0.0247 & 0.0186\\*
 &  & ~ & ~ & (\textbf{0.0000}) & (\textbf{0.0000}) & (0.0247) & (0.0186)\\*
\cmidrule{2-8}
 & \multirow{2}{*}{$100$} & \multirow{2}{*}{0.5264} & \multirow{2}{*}{0.4914} & 0.3262 & \textbf{0.0000} & 0.0485 & 0.0537\\*
 &  & ~ & ~ & (\textbf{0.0000}) & (\textbf{0.0000}) & (0.0485) & (0.0537)\\*
 \cmidrule{2-8}
 & \multirow{2}{*}{$1000$} & \multirow{2}{*}{0.5631} & \multirow{2}{*}{0.4990} & 0.3329 & \textbf{0.0000} & 0.0208 & 0.0225\\*
 &  & ~ & ~ & (\textbf{0.0000}) & (\textbf{0.0000}) & (0.0208) & (0.0225)\\
 \midrule
\multirow{6}{*}{General} & \multirow{2}{*}{$10$} & \multirow{2}{*}{0.2749} & \multirow{2}{*}{0.2092} & 0.2435 & 0.1656 & \textbf{0.0522} & 0.0548\\*
 &  & ~ & ~ & (0.0829) & (0.1729) & (0.0530) & (0.0548)\\*
\cmidrule{2-8}
 & \multirow{2}{*}{$100$} & \multirow{2}{*}{0.3816} & \multirow{2}{*}{0.2822} & 0.3092 & 0.0960 & 0.0461 & 0.0473\\*
 &  & ~ & ~ & (\textbf{0.0241}) & (0.0960) & (0.0461) & (0.0473)\\*
\cmidrule{2-8}
 & \multirow{2}{*}{$1000$} & \multirow{2}{*}{0.4170} & \multirow{2}{*}{0.3319} & 0.3259 & 0.0381 & 0.0205 & 0.0203\\*
 &  & ~ & ~ & (\textbf{0.0068}) & (0.0381) & (0.0205) & (0.0203)\\
\bottomrule
\end{tabular}

\end{table}

\begin{table}[H]
\centering
\caption{Well-supported approximation results for approximate NE solvers. We omit the publishing year in the algorithm name. Each datum in the table represents the average well-supported approximation of experiments on randomly generated games. The data in the parentheses is the average well-supported approximation before the mixing phase.}\label{tab:approxWSNE-of-approxNE-algo}
\begin{tabular}{cccccccc}
\toprule
\multicolumn{2}{c}{Scenario} & KPS-$0.75$ & DMP-$0.50$ & CDFFJS-$0.38$ & BBM-$0.36$ & TS-$0.3393$ & DFM-$1/3$\\
\midrule
\midrule
\multirow{6}{*}{Zero-Sum} & \multirow{2}{*}{$10$} & \multirow{2}{*}{0.7519} & \multirow{2}{*}{0.8560} & 0.7820 & \textbf{0.0000} & 0.1784 & 0.1538\\*
&  & ~ & ~ & (\textbf{0.0000}) & (\textbf{0.0000}) & (0.1784) & (0.1538)\\*
\cmidrule{2-8}
& \multirow{2}{*}{$100$} & \multirow{2}{*}{0.7962} & \multirow{2}{*}{0.9827} & 0.9652 & \textbf{0.0000} & 0.1855 & 0.1938\\*
&  & ~ & ~ & (\textbf{0.0000}) & (\textbf{0.0000}) & (0.1855) & (0.1938)\\*
\cmidrule{2-8}
& \multirow{2}{*}{$1000$} & \multirow{2}{*}{0.8406} & \multirow{2}{*}{0.9980} & 0.9977 & \textbf{0.0000} & 0.0909 & 0.0940\\*
&  & ~ & ~ & (0.0000) & (\textbf{0.0000}) & (0.0909) & (0.0940)\\
\midrule
\multirow{6}{*}{General} & \multirow{2}{*}{$10$} & \multirow{2}{*}{0.4876} & \multirow{2}{*}{0.4010} & 0.7233 & 0.3282 & 0.2884 & 0.3141\\*
&  & ~ & ~ & (\textbf{0.2383}) & (0.3283) & (0.2877) & (0.3141)\\*
\cmidrule{2-8}
& \multirow{2}{*}{$100$} & \multirow{2}{*}{0.5539} & \multirow{2}{*}{0.5161} & 0.9697 & 0.1763 & 0.2069 & 0.2117\\*
&  & ~ & ~ & (\textbf{0.1251}) & (0.1763) & (0.2069) & (0.2117)\\*
\cmidrule{2-8}
& \multirow{2}{*}{$1000$} & \multirow{2}{*}{0.6001} & \multirow{2}{*}{0.5798} & 0.9956 & 0.0736 & 0.0965 & 0.0926\\*
&  & ~ & ~ & (\textbf{0.0495}) & (0.0736) & (0.0965) & (0.0926)\\
\bottomrule
\end{tabular}

\end{table}

\begin{table}[H]
\centering
\caption{Approximation Results for Well-supported NE solvers. Each datum in the table represents the average well-supported approximation of experiments on randomly generated games. The data in the parentheses is the average approximation on the same output strategy profile.}\label{tab:results-of-approxWSNE-algo}
\begin{tabular}{cccccc}
\toprule
\multicolumn{2}{c}{Scenario} & KS07-$2/3$ & FGSS12-$0.6607$ & CDFFJS15-$0.6528$ & DFM22-$1/2$\\
\midrule
\midrule
\multirow{6}{*}{Zero-Sum} & \multirow{2}{*}{$10$} & 0.5022 & \textbf{0.0000} & \textbf{0.0000} & 0.1733\\*
 &  & (0.5022) & (\textbf{0.0000}) & (\textbf{0.0000}) & (0.0452)\\*
\cmidrule{2-6}
 & \multirow{2}{*}{$100$} & 0.5674 & \multirow{2}{*}{Timeout} & \textbf{0.0000} & 0.0374\\*
 &  & (0.5674) & ~ & (\textbf{0.0000}) & (0.0032)\\*
\cmidrule{2-6}
 & \multirow{2}{*}{$1000$} & 0.5826 & \multirow{2}{*}{Timeout} & \textbf{0.0000} & 0.0117\\*
 &  & (0.5826) & ~ & (\textbf{0.0000}) & (0.0004)\\
\midrule
\multirow{6}{*}{General} & \multirow{2}{*}{$10$} & 0.3740 & \textbf{0.0000} & 0.2383 & \multirow{2}{*}{Timeout}\\*
 &  & (0.3740) & (\textbf{0.0000}) & (0.0829) & ~\\*
\cmidrule{2-6}
 & \multirow{2}{*}{$100$} & 0.3661 &  \multirow{2}{*}{Timeout} & \textbf{0.1251} & \multirow{2}{*}{Timeout}\\*
 &  & (0.3661) &  & (\textbf{0.0241}) & ~\\*
\cmidrule{2-6}
 & \multirow{2}{*}{$1000$} & 0.4758 & \multirow{2}{*}{Timeout} & \textbf{0.0495} & \multirow{2}{*}{Timeout}\\*
 &  & (0.4758) & ~ & (\textbf{0.0068}) & ~\\
\bottomrule
\end{tabular}

\end{table}

\subsection{Learning dynamics on random games}\label{subsec:experiment-dynamics}

\begin{table}[H]
\centering
\caption{Approximation results for Learning Dynamics. Each datum in the table represents the average approximation of experiments for the average policy on randomly generated games. The data in the parentheses is the average approximation for the last iteration.}
\label{tab:ld_ne}
\begin{tabular}{ccccccc}
	\toprule
	\multicolumn{2}{c}{Scenario} & Fictitious Play & Hedge & MWU-exp & MWU-linear & Regret Matching\\
	\midrule
	\midrule
	\multirow{6}{*}{Zero-Sum} & \multirow{2}{*}{$10$} & \textbf{0.0016} & 0.0059 & 0.0018 & 0.0131 & 0.0019\\*
	&  & (0.6027) & (0.1175) & (0.6048) & (0.5695) & (0.3334)\\*
	\cmidrule{2-7}
	& \multirow{2}{*}{$100$} & 0.0047 & 0.0079 & 0.0047 & 0.0092 & \textbf{0.0033}\\*
	&  & (0.7235) & (0.0649) & (0.6836) & (0.6771) & (0.1342)\\*
	\cmidrule{2-7}
	& \multirow{2}{*}{$1000$} & 0.0094 & 0.0092 & 0.0090 & 0.0100 & \textbf{0.0042}\\*
	&  & (0.7645) & (0.0273) & (0.6577) & (0.5770) & (0.0577)\\
	\midrule
	\multirow{6}{*}{General} & \multirow{2}{*}{$10$} & 0.0129 & 0.0325 & 0.0181 & 0.0174 & \textbf{0.0118}\\*
	&  & (0.1668) & (0.1036) & (0.1624) & (0.1309) & (0.1146)\\*
	\cmidrule{2-7}
	& \multirow{2}{*}{$100$} & 0.0955 & 0.1012 & 0.0831 & \textbf{0.0594} & 0.0819\\*
	&  & (0.1618) & (0.1572) & (0.1127) & (0.1791) & (0.1463)\\*
	\cmidrule{2-7}
	& \multirow{2}{*}{$1000$} & 0.1366 & \textbf{0.1023} & 0.1262 & 0.1238 & 0.1115\\*
	&  & (0.1430) & (0.1159) & (0.1128) & (0.1244) & (0.1237)\\
	\bottomrule
\end{tabular}

\end{table}

\begin{table}[H]
\centering
\caption{Well-supported approximation results for Learning Dynamics. Each datum in the table represents the average well-supported approximation of experiments for the average policy on randomly generated games. The data in the parentheses is the average well-supported approximation for the last iteration.}
\label{tab:ld_wsne}
\begin{tabular}{ccccccc}
	\toprule
	\multicolumn{2}{c}{Scenario} & Fictitious Play & Hedge & MWU-exp & MWU-linear & Regret Matching\\
	\midrule
	\midrule
	\multirow{6}{*}{Zero-Sum} & \multirow{2}{*}{$10$} & \textbf{0.1530} & 0.2702 & 0.2720 & 0.2840 & 0.2722\\*
	&  & (0.6027) & (0.2340) & (0.6644) & (0.6297) & (0.4947)\\*
	\cmidrule{2-7}
	& \multirow{2}{*}{$100$} & \textbf{0.0939} & 0.1347 & 0.1373 & 0.1419 & 0.1397\\*
	&  & (0.7235) & (0.1657) & (0.8548) & (0.8557) & (0.2429)\\*
	\cmidrule{2-7}
	& \multirow{2}{*}{$1000$} & \textbf{0.0492} & 0.0526 & 0.0597 & 0.0612 & 0.0565\\*
	&  & (0.7645) & (0.0837) & (0.8787) & (0.9021) & (0.1129)\\
	\midrule
	\multirow{6}{*}{General} & \multirow{2}{*}{$10$} & 0.6591 & 0.7205 & 0.7695 & 0.7937 & 0.7887\\*
	&  & (0.1668) & (0.1702) & (0.1717) & (0.1428) & (\textbf{0.1155})\\*
	\cmidrule{2-7}
	& \multirow{2}{*}{$100$} & 0.7225 & 0.6741 & 0.8215 & 0.8209 & 0.8239\\*
	&  & (0.1618) & (0.3599) & (\textbf{0.1162}) & (0.1791) & (0.1469)\\*
	\cmidrule{2-7}
	& \multirow{2}{*}{$1000$} & 0.6337 & 0.3541 & 0.7646 & 0.7945 & 0.8120\\*
	&  & (0.1430) & (0.5689) & (\textbf{0.1139}) & (0.1244) & (0.1237)\\
	\bottomrule
\end{tabular}
\end{table}

\subsection{Algorithms with specific approximation guarantees on bimatrix GAMUT games}

\begin{longtable}{cccccccc}
\caption{Approximation results for approximate NE solvers on GAMUT. We omit the publishing year in the algorithm name. Each data point in the table represents the average approximation of experiments on randomly generated games. The data in the parentheses is the average approximation before the mixing phase.}\\
\toprule
\multicolumn{2}{c}{Scenario} & KPS-$0.75$ & DMP-$0.50$ & CDFFJS-$0.38$ & BBM-$0.36$ & TS07-$0.3393$ & DFM-$1/3$\\
\midrule
\midrule
\endfirsthead\multicolumn{8}{l}{\tablename\ \thetable\ -- (\textit{Continued})} \\
\toprule
\multicolumn{2}{c}{Scenario} & KPS-$0.75$ & DMP-$0.50$ & CDFFJS-$0.38$ & BBM-$0.36$ & TS-$0.3393$ & DFM-$1/3$\\
\midrule
\midrule
\endhead
\bottomrule
\multicolumn{8}{r}{\textit{Continued on next page}} \\
\endfoot
\bottomrule
\endlastfoot
\multirow{6}{*}{BO} & \multirow{2}{*}{$10$} & \multirow{2}{*}{0.3507} & \multirow{2}{*}{0.2674} & 0.0450 & 0.0493 & \textbf{0.0064} & 0.0144\\*
 &  & ~ & ~ & (0.0432) & (0.0523) & (0.0138) & (0.0144)\\*
\cmidrule{2-8}
 & \multirow{2}{*}{$100$} & \multirow{2}{*}{0.4395} & \multirow{2}{*}{0.3036} & 0.0039 & 0.0039 & \textbf{0.0000} & 0.0000\\*
 &  & ~ & ~ & (0.0039) & (0.0039) & (\textbf{0.0000}) & (0.0000)\\*
\cmidrule{2-8}
 & \multirow{2}{*}{$1000$} & \multirow{2}{*}{0.4990} & \multirow{2}{*}{0.3080} & 0.0002 & 0.0002 & \textbf{0.0000} & \textbf{0.0000}\\*
 &  & ~ & ~ & (0.0002) & (0.0002) & (\textbf{0.0000}) & (\textbf{0.0000})\\
\cmidrule{1-8}
\multirow{6}{*}{CD} & \multirow{2}{*}{$10$} & \multirow{2}{*}{\textbf{0.0077}} & \multirow{2}{*}{0.1033} & 0.0206 & 0.0305 & 0.0281 & 0.0336\\*
 &  & ~ & ~ & (0.0357) & (0.0365) & (0.0322) & (0.0336)\\*
\cmidrule{2-8}
 & \multirow{2}{*}{$100$} & \multirow{2}{*}{0.0008} & \multirow{2}{*}{0.1873} & 0.0008 & \textbf{0.0006} & 0.0181 & 0.0201\\*
 &  & ~ & ~ & (0.0008) & (\textbf{0.0006}) & (0.0190) & (0.0201)\\*
\cmidrule{2-8}
 & \multirow{2}{*}{$1000$} & \multirow{2}{*}{\textbf{0.0000}} & \multirow{2}{*}{\textbf{0.0000}} & \textbf{0.0000} & 0.0001 & 0.0003 & 0.0004\\*
 &  & ~ & ~ & (\textbf{0.0000}) & (0.0001) & (0.0003) & (0.0004)\\
\cmidrule{1-8}
\multirow{6}{*}{CG} & \multirow{2}{*}{$10$} & \multirow{2}{*}{0.1512} & \multirow{2}{*}{0.0868} & 0.1651 & 0.1432 & 0.0461 & \textbf{0.0455}\\*
 &  & ~ & ~ & (0.0555) & (0.1432) & (0.0469) & (\textbf{0.0455})\\*
 \cmidrule{2-8}
 & \multirow{2}{*}{$100$} & \multirow{2}{*}{0.1166} & \multirow{2}{*}{0.1068} & 0.2009 & 0.0473 & 0.0211 & 0.0201\\*
 &  & ~ & ~ & (\textbf{0.0117}) & (0.0473) & (0.0211) & (0.0201)\\*
 \cmidrule{2-8}
 & \multirow{2}{*}{$1000$} & \multirow{2}{*}{0.1046} & \multirow{2}{*}{0.0908} & 0.2077 & 0.0158 & 0.0073 & 0.0074\\*
 &  & ~ & ~ & (\textbf{0.0028}) & (0.0158) & (0.0073) & (0.0074)\\
 \cmidrule{1-8}
\multirow{6}{*}{GTD} & \multirow{2}{*}{$10$} & \multirow{2}{*}{0.6079} & \multirow{2}{*}{0.2106} & 0.0176 & 0.0954 & 0.0073 & 0.0149\\*
 &  & ~ & ~ & (\textbf{0.0000}) & (0.5683) & (0.0143) & (0.0149)\\*
 \cmidrule{2-8}
 & \multirow{2}{*}{$100$} & \multirow{2}{*}{0.5976} & \multirow{2}{*}{0.1951} &  \multirow{2}{*}{Precision Error} & 0.0956 & \textbf{0.0012} & 0.0020\\*
 &  & ~ & ~ &  & (0.6098) & (0.0015) & (0.0020)\\*
 \cmidrule{2-8}
 & \multirow{2}{*}{$1000$} & \multirow{2}{*}{0.5806} & \multirow{2}{*}{0.1611} &  \multirow{2}{*}{Precision Error} & 0.0603 & 0.0001 & \textbf{0.0001}\\*
 &  & ~ & ~ &  & (0.6777) & (0.0002) & (\textbf{0.0001})\\
 \cmidrule{1-8}
\multirow{6}{*}{GTTA} & \multirow{2}{*}{$10$} & \multirow{2}{*}{0.5000} & \multirow{2}{*}{0.4062} & \textbf{0.0000} & \textbf{0.0000} & \textbf{0.0000} & \textbf{0.0000}\\*
 &  & ~ & ~ & (\textbf{0.0000}) & (\textbf{0.0000}) & (\textbf{0.0000}) & (\textbf{0.0000})\\*
 \cmidrule{2-8}
 & \multirow{2}{*}{$100$} & \multirow{2}{*}{0.5000} & \multirow{2}{*}{0.5000} & \textbf{0.0000} & \textbf{0.0000} & \textbf{0.0000} & \textbf{0.0000}\\*
 &  & ~ & ~ & (\textbf{0.0000}) & (\textbf{0.0000}) & (\textbf{0.0000}) & (\textbf{0.0000})\\*
 \cmidrule{2-8}
 & \multirow{2}{*}{$1000$} & \multirow{2}{*}{0.5000} & \multirow{2}{*}{0.4750} & \textbf{0.0000} & \textbf{0.0000} & \textbf{0.0000} & \textbf{0.0000}\\*
 &  & ~ & ~ & (\textbf{0.0000}) & (\textbf{0.0000}) & (\textbf{0.0000}) & (\textbf{0.0000})\\
 \cmidrule{1-8}
\multirow{6}{*}{LG} & \multirow{2}{*}{$10$} & \multirow{2}{*}{0.0023} & \multirow{2}{*}{0.0018} & \textbf{0.0010} & 0.0012 & 0.0027 & 0.0029\\*
 &  & ~ & ~ & (0.0012) & (0.0012) & (0.0027) & (0.0029)\\*
 \cmidrule{2-8}
 & \multirow{2}{*}{$100$} & \multirow{2}{*}{0.0250} & \multirow{2}{*}{\textbf{0.0008}} & 0.0061 & 0.0110 & 0.0056 & 0.0059\\*
 &  & ~ & ~ & (0.0068) & (0.0110) & (0.0056) & (0.0059)\\*
 \cmidrule{2-8}
 & \multirow{2}{*}{$1000$} & \multirow{2}{*}{0.0266} & \multirow{2}{*}{0.0130} & 0.0075 & 0.0351 & 0.0217 & \textbf{0.0024}\\*
 &  & ~ & ~ & (0.0082) & (0.0351) & (0.0217) & (\textbf{0.0024})\\
 \cmidrule{1-8}
\multirow{6}{*}{MEG} & \multirow{2}{*}{$10$} & \multirow{2}{*}{\textbf{0.0000}} & \multirow{2}{*}{\textbf{0.0000}} & \textbf{0.0000} & \textbf{0.0000} & 0.0070 & 0.0217\\*
 &  & ~ & ~ & (\textbf{0.0000}) & (\textbf{0.0000}) & (0.0080) & (0.0217)\\*
 \cmidrule{2-8}
 & \multirow{2}{*}{$100$} & \multirow{2}{*}{\textbf{0.0000}} & \multirow{2}{*}{\textbf{0.0000}} & 0.0072 & \textbf{0.0000} & \textbf{0.0000} & \textbf{0.0000}\\*
 &  & ~ & ~ & (\textbf{0.0000}) & (\textbf{0.0000}) & (\textbf{0.0000}) & (\textbf{0.0000})\\*
 \cmidrule{2-8}
 & \multirow{2}{*}{$1000$} & \multirow{2}{*}{\textbf{0.0000}} & \multirow{2}{*}{\textbf{0.0000}} & \textbf{0.0000} & \textbf{0.0000} & \textbf{0.0000} & \textbf{0.0000}\\*
 &  & ~ & ~ & (\textbf{0.0000}) & (\textbf{0.0000}) & (\textbf{0.0000}) & (\textbf{0.0000})\\
 \cmidrule{1-8}
\multirow{6}{*}{TD} & \multirow{2}{*}{$10$} & \multirow{2}{*}{0.1376} & \multirow{2}{*}{0.2150} & \textbf{0.0000} & \textbf{0.0000} & \textbf{0.0000} & \textbf{0.0000}\\*
 &  & ~ & ~ & (0.4864) & (\textbf{0.0000}) & (\textbf{0.0000}) & (\textbf{0.0000})\\*
 \cmidrule{2-8}
 & \multirow{2}{*}{$100$} & \multirow{2}{*}{0.1085} & \multirow{2}{*}{0.2170} & \textbf{0.0000} & \textbf{0.0000} & \textbf{0.0000} & \textbf{0.0000}\\*
 &  & ~ & ~ & (0.4281) & (\textbf{0.0000}) & (\textbf{0.0000}) & (\textbf{0.0000})\\*
 \cmidrule{2-8}
 & \multirow{2}{*}{$1000$} & \multirow{2}{*}{0.1058} & \multirow{2}{*}{0.2117} & 0.0537 & \textbf{0.0000} & \textbf{0.0000} & \textbf{0.0000}\\*
 &  & ~ & ~ & (0.4228) & (\textbf{0.0000}) & (\textbf{0.0000}) & (\textbf{0.0000})\\
 \cmidrule{1-8}
\multirow{6}{*}{WOA} & \multirow{2}{*}{$10$} & \multirow{2}{*}{0.1409} & \multirow{2}{*}{0.1144} & \textbf{0.0000} & \textbf{0.0000} & 0.0039 & 0.0074\\*
 &  & ~ & ~ & (\textbf{0.0000}) & (\textbf{0.0000}) & (0.0078) & (0.0074)\\*
 \cmidrule{2-8}
 & \multirow{2}{*}{$100$} & \multirow{2}{*}{0.0218} & \multirow{2}{*}{0.0173} & \textbf{0.0000} & \textbf{0.0000} & 0.0006 & 0.0006\\*
 &  & ~ & ~ & (\textbf{0.0000}) & (\textbf{0.0000}) & (0.0006) & (0.0006)\\*
 \cmidrule{2-8}
 & \multirow{2}{*}{$1000$} &
 \multirow{2}{*}{0.0025} & \multirow{2}{*}{0.0021} & \textbf{0.0000} & \textbf{0.0000} & 0.0101 & 0.0101\\*
 &  & ~ & ~ & (\textbf{0.0000}) & (\textbf{0.0000}) & (0.0101) & (0.0101)\\
\end{longtable}

\begin{longtable}{cccccccc}
\caption{Well-supported approximation results for approximate NE solvers. We omit the publishing year in the algorithm name. Each datum in the table represents the average well-supported approximation of experiments on randomly generated games. The data in the parentheses is average well-supported approximation before the mixing phase.}\\
\toprule
\multicolumn{2}{c}{Scenario} & KPS-$0.75$ & DMP-$0.50$ & CDFFJS-$0.38$ & BBM-$0.36$ & TS-$0.3393$ & DFM-$1/3$\\
\midrule
\midrule
\endfirsthead\multicolumn{8}{l}{\tablename\ \thetable\ -- (\textit{Continued})} \\
\toprule
\multicolumn{2}{c}{Scenario} & KPS-$0.75$ & DMP-$0.50$ & CDFFJS-$0.38$ & BBM-$0.36$ & TS-$0.3393$ & DFM-$1/3$\\
\midrule
\midrule
\endhead
\bottomrule
\multicolumn{8}{r}{\textit{Continued on next page}} \\
\endfoot
\bottomrule
\endlastfoot
\multirow{6}{*}{BO} & \multirow{2}{*}{$10$} & \multirow{2}{*}{0.5644} & \multirow{2}{*}{0.4530} & 0.0568 & 0.0510 & 0.0860 & 0.1294\\*
 &  & ~ & ~ & (\textbf{0.0439}) & (0.0523) & (0.1346) & (0.1294)\\*
 \cmidrule{2-8}
 & \multirow{2}{*}{$100$} & \multirow{2}{*}{0.6810} & \multirow{2}{*}{0.5967} & \textbf{0.0039} & \textbf{0.0039} & 0.1435 & 0.1468\\*
 &  & ~ & ~ & (\textbf{0.0039}) & (\textbf{0.0039}) & (0.1435) & (0.1468)\\*
 \cmidrule{2-8}
 & \multirow{2}{*}{$1000$} & \multirow{2}{*}{0.7486} & \multirow{2}{*}{0.6144} & \textbf{0.0002} & \textbf{0.0002} & 0.1274 & 0.1277\\*
 &  & ~ & ~ & (\textbf{0.0002}) & (\textbf{0.0002}) & (0.1274) & (0.1277)\\
 \midrule
\multirow{6}{*}{CD} & \multirow{2}{*}{$10$} & \multirow{2}{*}{\textbf{0.0153}} & \multirow{2}{*}{0.2066} & 0.0206 & 0.0379 & 0.0585 & 0.0705\\*
 &  & ~ & ~ & (0.0357) & (0.0379) & (0.0738) & (0.0705)\\*
 \cmidrule{2-8}
 & \multirow{2}{*}{$100$} & \multirow{2}{*}{0.0015} & \multirow{2}{*}{0.3747} & 0.0008 & \textbf{0.0006} & 0.0337 & 0.0359\\*
 &  & ~ & ~ & (0.0008) & (\textbf{0.0006}) & (0.0349) & (0.0359)\\*
 \cmidrule{2-8}
 & \multirow{2}{*}{$1000$} & \multirow{2}{*}{\textbf{0.0000}} & \multirow{2}{*}{\textbf{0.0000}} & \textbf{0.0000} & 0.0001 & 0.0003 & 0.0004\\*
 &  & ~ & ~ & (\textbf{0.0000}) & (0.0001) & (0.0003) & (0.0004)\\
 \midrule
\multirow{6}{*}{CG} & \multirow{2}{*}{$10$} & \multirow{2}{*}{0.2799} & \multirow{2}{*}{0.1636} & 0.5504 & 0.2608 & 0.2092 & 0.1959\\*
 &  & ~ & ~ & (\textbf{0.1611}) & (0.2608) & (0.2117) & (0.1959)\\*
 \cmidrule{2-8}
 & \multirow{2}{*}{$100$} & \multirow{2}{*}{0.2194} & \multirow{2}{*}{0.2121} & 0.5812 & 0.0894 & 0.0985 & 0.0988\\*
 &  & ~ & ~ & (\textbf{0.0593}) & (0.0894) & (0.0985) & (0.0988)\\*
 \cmidrule{2-8}
 & \multirow{2}{*}{$1000$} & \multirow{2}{*}{0.1824} & \multirow{2}{*}{0.1816} & 0.6333 & 0.0314 & 0.0366 & 0.0351\\*
 &  & ~ & ~ & (\textbf{0.0187}) & (0.0314) & (0.0366) & (0.0351)\\
 \midrule
\multirow{6}{*}{GTD} & \multirow{2}{*}{$10$} & \multirow{2}{*}{0.7159} & \multirow{2}{*}{0.4130} & 0.5177 & 0.1004 & 0.3072 & 0.4786\\*
 &  & ~ & ~ & (\textbf{0.0000}) & (0.5683) & (0.4423) & (0.4786)\\*
 \cmidrule{2-8}
 & \multirow{2}{*}{$100$} & \multirow{2}{*}{0.6951} & \multirow{2}{*}{0.3846} &  \multirow{2}{*}{Precision Error} & \textbf{0.0990} & 0.2837 & 0.4000\\*
 &  & ~ & ~ &  & (0.6098) & (0.4436) & (0.4000)\\*
 \cmidrule{2-8}
 & \multirow{2}{*}{$1000$} & \multirow{2}{*}{0.6611} & \multirow{2}{*}{0.3223} &  \multirow{2}{*}{Precision Error} & \textbf{0.0660} & 0.3084 & 0.4437\\*
 &  & ~ & ~ &  & (0.6777) & (0.4735) & (0.4437)\\
 \midrule
\multirow{6}{*}{GTTA} & \multirow{2}{*}{$10$} & \multirow{2}{*}{0.7500} & \multirow{2}{*}{0.8125} & \textbf{0.0000} & \textbf{0.0000} & \textbf{0.0000} & \textbf{0.0000}\\*
 &  & ~ & ~ & (\textbf{0.0000}) & (\textbf{0.0000}) & (\textbf{0.0000}) & (\textbf{0.0000})\\*
 \cmidrule{2-8}
 & \multirow{2}{*}{$100$} & \multirow{2}{*}{0.7500} & \multirow{2}{*}{1.0000} & \textbf{0.0000} & \textbf{0.0000} & \textbf{0.0000} & \textbf{0.0000}\\*
 &  & ~ & ~ & (\textbf{0.0000}) & (\textbf{0.0000}) & (\textbf{0.0000}) & (\textbf{0.0000})\\*
 \cmidrule{2-8}
 & \multirow{2}{*}{$1000$} & \multirow{2}{*}{0.7500} & \multirow{2}{*}{0.9500} & \textbf{0.0000} & \textbf{0.0000} & \textbf{0.0000} & \textbf{0.0000}\\*
 &  & ~ & ~ & (\textbf{0.0000}) & (\textbf{0.0000}) & (\textbf{0.0000}) & (\textbf{0.0000})\\
 \midrule
\multirow{6}{*}{LG} & \multirow{2}{*}{$10$} & \multirow{2}{*}{0.0047} & \multirow{2}{*}{0.0036} & \textbf{0.0010} & 0.0012 & 0.0033 & 0.0059\\*
 &  & ~ & ~ & (0.0012) & (0.0012) & (0.0059) & (0.0059)\\*
 \cmidrule{2-8}
 & \multirow{2}{*}{$100$} & \multirow{2}{*}{0.0500} & \multirow{2}{*}{\textbf{0.0015}} & 0.0061 & 0.0110 & 0.0118 & 0.0115\\*
 &  & ~ & ~ & (0.0068) & (0.0110) & (0.0118) & (0.0115)\\*
 \cmidrule{2-8}
 & \multirow{2}{*}{$1000$} & \multirow{2}{*}{0.0532} & \multirow{2}{*}{0.0171} & 0.0075 & 0.0351 & 0.0597 & \textbf{0.0049}\\*
 &  & ~ & ~ & (0.0082) & (0.0351) & (0.0597) & (\textbf{0.0049})\\
 \midrule
\multirow{6}{*}{MEG} & \multirow{2}{*}{$10$} & \multirow{2}{*}{\textbf{0.0000}} & \multirow{2}{*}{\textbf{0.0000}} & \textbf{0.0000} & \textbf{0.0000} & 0.0321 & 0.1173\\*
 &  & ~ & ~ & (\textbf{0.0000}) & (\textbf{0.0000}) & (0.0419) & (0.1173)\\*
 \cmidrule{2-8}
 & \multirow{2}{*}{$100$} & \multirow{2}{*}{\textbf{0.0000}} & \multirow{2}{*}{\textbf{0.0000}} & 0.0232 & \textbf{0.0000} & \textbf{0.0000} & \textbf{0.0000}\\*
 &  & ~ & ~ & (\textbf{0.0000}) & (\textbf{0.0000}) & (\textbf{0.0000}) & (\textbf{0.0000})\\*
 \cmidrule{2-8}
 & \multirow{2}{*}{$1000$} & \multirow{2}{*}{\textbf{0.0000}} & \multirow{2}{*}{\textbf{0.0000}} & \textbf{0.0000} & \textbf{0.0000} & \textbf{0.0000} & \textbf{0.0000}\\*
 &  & ~ & ~ & (\textbf{0.0000}) & (\textbf{0.0000}) & (\textbf{0.0000}) & (\textbf{0.0000})\\
 \midrule
\multirow{6}{*}{TD} & \multirow{2}{*}{$10$} & \multirow{2}{*}{0.2753} & \multirow{2}{*}{0.4299} & \textbf{0.0000} & \textbf{0.0000} & \textbf{0.0000} & \textbf{0.0000}\\*
 &  & ~ & ~ & (0.4864) & (\textbf{0.0000}) & (\textbf{0.0000}) & (\textbf{0.0000})\\*
 \cmidrule{2-8}
 & \multirow{2}{*}{$100$} & \multirow{2}{*}{0.2170} & \multirow{2}{*}{0.4339} & \textbf{0.0000} & \textbf{0.0000} & \textbf{0.0000} & \textbf{0.0000}\\*
 &  & ~ & ~ & (0.4281) & (\textbf{0.0000}) & (\textbf{0.0000}) & (\textbf{0.0000})\\*
 \cmidrule{2-8}
 & \multirow{2}{*}{$1000$} & \multirow{2}{*}{0.2117} & \multirow{2}{*}{0.4234} & 0.0847 & \textbf{0.0000} & \textbf{0.0000} & \textbf{0.0000}\\*
 &  & ~ & ~ & (0.4228) & (\textbf{0.0000}) & (\textbf{0.0000}) & (\textbf{0.0000})\\
 \midrule
\multirow{6}{*}{WOA} & \multirow{2}{*}{$10$} & \multirow{2}{*}{0.2818} & \multirow{2}{*}{0.2289} & \textbf{0.0000} & \textbf{0.0000} & 0.1340 & 0.1856\\*
 &  & ~ & ~ & (\textbf{0.0000}) & (\textbf{0.0000}) & (0.1676) & (0.1856)\\*
 \cmidrule{2-8}
 & \multirow{2}{*}{$100$} & \multirow{2}{*}{0.0435} & \multirow{2}{*}{0.0346} & \textbf{0.0000} & \textbf{0.0000} & 0.0392 & 0.0393\\*
 &  & ~ & ~ & (\textbf{0.0000}) & (\textbf{0.0000}) & (0.0392) & (0.0393)\\*
 \cmidrule{2-8}
 & \multirow{2}{*}{$1000$} & \multirow{2}{*}{0.0050} & \multirow{2}{*}{0.0042} & \textbf{0.0000} & \textbf{0.0000} & 0.1546 & 0.1538\\*
 &  & ~ & ~ & (\textbf{0.0000}) & (\textbf{0.0000}) & (0.1546) & (0.1538)\\
\end{longtable}

\begin{longtable}{cccccc}
\caption{Approximation Results for well-supported NE solvers on GAMUT. Each datum in the table represents the average well-supported approximation of experiments on randomly generated games. The data in the parentheses is average approximation on the same output strategy profile.}\\
\toprule
\multicolumn{2}{c}{Scenario} & KS07-$2/3$ & FGSS12-$0.6607$ & CDFFJS15-$0.6528$ & DFM22-$1/2$\\
\midrule
\midrule
\endfirsthead\multicolumn{6}{l}{\tablename\ \thetable\ -- (\textit{Continued})} \\
\toprule
\multicolumn{2}{c}{Scenario} & KS07-$2/3$ & FGSS12-$0.6607$ & CDFFJS15-$0.6528$ & DFM22-$1/2$\\
\midrule
\midrule
\endhead
\bottomrule
\multicolumn{6}{r}{\textit{Continued on next page}} \\
\endfoot
\bottomrule
\endlastfoot
\multirow{6}{*}{BO} & \multirow{2}{*}{$10$} & 0.3460 & \textbf{0.0000} & 0.0497 &  \multirow{2}{*}{Timeout}\\*
 &  & (0.3460) & (\textbf{0.0000}) & (0.0447) & \\*
 \cmidrule{2-6}
 & \multirow{2}{*}{$100$} & 0.3350 & \textbf{0.0000} & 0.0040 & 0.0040\\*
 &  & (0.3350) & (\textbf{0.0000}) & (0.0040) & (0.0040)\\*
 \cmidrule{2-6}
 & \multirow{2}{*}{$1000$} & 0.3335 & \textbf{0.0000} & 0.0001 & 0.0001\\*
 &  & (0.3335) & (\textbf{0.0000}) & (0.0001) & (0.0001)\\
 \midrule
\multirow{6}{*}{CD} & \multirow{2}{*}{$10$} & 0.1244 & \textbf{0.0000} & 0.0357 &  \multirow{2}{*}{Timeout}\\*
 &  & (0.1244) & (\textbf{0.0000}) & (0.0357) & \\*
 \cmidrule{2-6}
 & \multirow{2}{*}{$100$} & 0.2164 & \textbf{0.0000} & 0.0008 &  \multirow{2}{*}{Timeout}\\*
 &  & (0.2164) & (\textbf{0.0000}) & (0.0008) & \\*
 \cmidrule{2-6}
 & \multirow{2}{*}{$1000$} & 0.0623 & \textbf{0.0000} & \textbf{0.0000} &  \multirow{2}{*}{Timeout}\\*
 &  & (0.0623) & (\textbf{0.0000}) & (\textbf{0.0000}) & \\
 \midrule
\multirow{6}{*}{CG} & \multirow{2}{*}{$10$} & 0.2894 & \textbf{0.0000} & 0.1611 & \multirow{2}{*}{Timeout}\\*
 &  & (0.2894) & (\textbf{0.0000}) & (0.0555) & ~\\*
 \cmidrule{2-6}
 & \multirow{2}{*}{$100$} & 0.3582 & \textbf{0.0000} & 0.0593 & \multirow{2}{*}{Timeout}\\*
 &  & (0.3582) & (\textbf{0.0000}) & (0.0117) & ~\\*
 \cmidrule{2-6}
 & \multirow{2}{*}{$1000$} & 0.3773 & \textbf{0.0000} & 0.0187 & \multirow{2}{*}{Timeout}\\*
 &  & (0.3773) & (\textbf{0.0000}) & (0.0028) & ~\\
 \midrule
\multirow{6}{*}{GTD} & \multirow{2}{*}{$10$} & 0.2475 & \textbf{0.0000} & 0.0039 &  \multirow{2}{*}{Timeout}\\*
 &  & (0.2475) & (\textbf{0.0000}) & (0.0037) & \\*
 \cmidrule{2-6}
 & \multirow{2}{*}{$100$} & 0.2773 & \textbf{0.0000} &  \multirow{2}{*}{Precision Error} &  \multirow{2}{*}{Timeout}\\*
 &  & (0.2773) & (\textbf{0.0000}) &  & \\*
 \cmidrule{2-6}
 & \multirow{2}{*}{$1000$} & 0.2235 & \textbf{0.0000} &  \multirow{2}{*}{Precision Error} &  \multirow{2}{*}{Timeout}\\*
 &  & (0.2235) & (\textbf{0.0000}) &  & \\
 \midrule
\multirow{6}{*}{GTTA} & \multirow{2}{*}{$10$} & 0.5000 & \textbf{0.0000} & \textbf{0.0000} & \textbf{0.0000}\\*
 &  & (0.5000) & (\textbf{0.0000}) & (\textbf{0.0000}) & (\textbf{0.0000})\\*
 \cmidrule{2-6}
 & \multirow{2}{*}{$100$} & 0.5000 & \textbf{0.0000} & \textbf{0.0000} & \textbf{0.0000}\\*
 &  & (0.5000) & (\textbf{0.0000}) & (\textbf{0.0000}) & (\textbf{0.0000})\\*
 \cmidrule{2-6}
 & \multirow{2}{*}{$1000$} & 0.5000 & \textbf{0.0000} & \textbf{0.0000} & \textbf{0.0000}\\*
 &  & (0.5000) & (\textbf{0.0000}) & (\textbf{0.0000}) & (\textbf{0.0000})\\
 \midrule
\multirow{6}{*}{LG} & \multirow{2}{*}{$10$} & 0.0059 & \textbf{0.0000} & 0.0012 & 0.0013\\*
 &  & (0.0059) & (\textbf{0.0000}) & (0.0012) & (0.0013)\\*
 \cmidrule{2-6}
 & \multirow{2}{*}{$100$} & 0.0074 & \textbf{0.0000} & 0.0068 & 0.0111\\*
 &  & (0.0074) & (\textbf{0.0000}) & (0.0068) & (0.0111)\\*
 \cmidrule{2-6}
 & \multirow{2}{*}{$1000$} & 0.0523 & \textbf{0.0000} & 0.0082 & 0.0082\\*
 &  & (0.0523) & (\textbf{0.0000}) & (0.0082) & (0.0082)\\
 \midrule
\multirow{6}{*}{MEG} & \multirow{2}{*}{$10$} & \textbf{0.0000} & \textbf{0.0000} & \textbf{0.0000} &  \multirow{2}{*}{Timeout}\\*
 &  & (\textbf{0.0000}) & (\textbf{0.0000}) & (\textbf{0.0000}) & \\*
 \cmidrule{2-6}
 & \multirow{2}{*}{$100$} & \textbf{0.0000} & \textbf{0.0000} & 0.0439 &  \multirow{2}{*}{Timeout}\\*
 &  & (\textbf{0.0000}) & (\textbf{0.0000}) & (0.0221) & \\*
 \cmidrule{2-6}
 & \multirow{2}{*}{$1000$} & \textbf{0.0000} & \textbf{0.0000} & 0.0901 &  \multirow{2}{*}{Timeout}\\*
 &  & (\textbf{0.0000}) & (\textbf{0.0000}) & (0.0422) & \\
 \midrule
\multirow{6}{*}{TD} & \multirow{2}{*}{$10$} & 0.4864 & \textbf{0.0000} & 0.4864 & 0.4864\\*
 &  & (0.4864) & (\textbf{0.0000}) & (0.4864) & (0.4864)\\*
 \cmidrule{2-6}
 & \multirow{2}{*}{$100$} & 0.4281 & \textbf{0.0000} & 0.4281 & 0.4281\\*
 &  & (0.4281) & (\textbf{0.0000}) & (0.4281) & (0.4281)\\*
 \cmidrule{2-6}
 & \multirow{2}{*}{$1000$} & 0.4228 & \textbf{0.0000} & 0.4228 & 0.4228\\*
 &  & (0.4228) & (\textbf{0.0000}) & (0.4228) & (0.4228)\\
 \midrule
\multirow{6}{*}{WOA} & \multirow{2}{*}{$10$} & \textbf{0.0000} & \textbf{0.0000} & \textbf{0.0000} &  \multirow{2}{*}{Timeout}\\*
 &  & (\textbf{0.0000}) & (\textbf{0.0000}) & (\textbf{0.0000}) & \\*
 \cmidrule{2-6}
 & \multirow{2}{*}{$100$} & \textbf{0.0000} & \textbf{0.0000} & \textbf{0.0000} &  \multirow{2}{*}{Timeout}\\*
 &  & (\textbf{0.0000}) & (\textbf{0.0000}) & (\textbf{0.0000}) & \\*
 \cmidrule{2-6}
 & \multirow{2}{*}{$1000$} & \textbf{0.0000} & \textbf{0.0000} & \textbf{0.0000} &  \multirow{2}{*}{Timeout}\\*
 &  & (\textbf{0.0000}) & (\textbf{0.0000}) & (\textbf{0.0000}) & \\

\end{longtable}

\subsection{Learning dynamics on bimatrix GAMUT games}
\label{sec:ld_gamut}

\begin{longtable}{ccccccc}
\caption{Approximation Results for Learning Dynamics on GAMUT. Each datum in the table represents the average approximation of experiments for the average policy on randomly generated games. The data in the parentheses is the average approximation for the last iteration.}\\
	\toprule
	\multicolumn{2}{c}{Scenario} & Fictitious Play & Hedge & MWU-exp & MWU-linear & Regret Matching\\
	\midrule
	\midrule
	\endfirsthead\multicolumn{7}{l}{\tablename\ \thetable\ -- (\textit{Continued})} \\
	\toprule
	\multicolumn{2}{c}{Scenario} & Fictitious Play & Hedge & MWU-exp & MWU-linear & Regret Matching\\
	\midrule
	\midrule
	\endhead
	\bottomrule
	\multicolumn{7}{r}{\textit{Continued on next page}} \\
	\endfoot
	\bottomrule
	\endlastfoot
	\multirow{6}{*}{BO} & \multirow{2}{*}{$10$} & 0.0000 & 0.0085 & 0.0001 & 0.0001 & 0.0000\\*
	&  & (0.0051) & (0.0000) & (\textbf{0.0000}) & (\textbf{0.0000}) & (\textbf{0.0000})\\*
	\cmidrule{2-7}
	& \multirow{2}{*}{$100$} & 0.0000 & 0.0060 & 0.0001 & 0.0001 & 0.0000\\*
	&  & (\textbf{0.0000}) & (0.0008) & (\textbf{0.0000}) & (\textbf{0.0000}) & (\textbf{0.0000})\\*
	\cmidrule{2-7}
	& \multirow{2}{*}{$1000$} & 0.0000 & 0.0107 & 0.0000 & 0.0000 & 0.0001\\*
	&  & (\textbf{0.0000}) & (0.0014) & (0.0000) & (0.0000) & (0.0000)\\
	\midrule
	\multirow{6}{*}{CD} & \multirow{2}{*}{$10$} & 0.0000 & 0.0075 & 0.0001 & 0.0001 & 0.0000\\*
	&  & (\textbf{0.0000}) & (0.0006) & (0.0000) & (0.0000) & (\textbf{-0.0000})\\*
	\cmidrule{2-7}
	& \multirow{2}{*}{$100$} & 0.0000 & 0.0611 & 0.0006 & 0.0006 & 0.0000\\*
	&  & (\textbf{0.0000}) & (0.0005) & (\textbf{0.0000}) & (\textbf{0.0000}) & (\textbf{0.0000})\\*
	\cmidrule{2-7}
	& \multirow{2}{*}{$1000$} & 0.0000 & 0.0006 & 0.0010 & 0.0008 & 0.0000\\*
	&  & (\textbf{0.0000}) & (0.0012) & (\textbf{0.0000}) & (\textbf{0.0000}) & (\textbf{0.0000})\\
	\midrule
	\multirow{6}{*}{CG} & \multirow{2}{*}{$10$} & \textbf{0.0006} & 0.0302 & 0.0075 & 0.0072 & 0.0012\\*
	&  & (0.0317) & (0.0208) & (0.0238) & (0.0413) & (0.0291)\\*
	\cmidrule{2-7}
	& \multirow{2}{*}{$100$} & \textbf{0.0093} & 0.0792 & 0.0157 & 0.0209 & 0.0125\\*
	&  & (0.0459) & (0.0736) & (0.0756) & (0.1041) & (0.0599)\\*
	\cmidrule{2-7}
	& \multirow{2}{*}{$1000$} & 0.0420 & \textbf{0.0128} & 0.0307 & 0.0371 & 0.0423\\*
	&  & (0.1125) & (0.0227) & (0.1097) & (0.1638) & (0.0816)\\
	\midrule
	\multirow{6}{*}{GTD} & \multirow{2}{*}{$10$} & 0.0000 & 0.0188 & 0.0001 & 0.0001 & 0.0000\\*
	&  & (0.0164) & (0.0003) & (\textbf{0.0000}) & (\textbf{0.0000}) & (\textbf{0.0000})\\*
	\cmidrule{2-7}
	& \multirow{2}{*}{$100$} & 0.0000 & 0.0302 & 0.0002 & 0.0002 & 0.0001\\*
	&  & (\textbf{0.0000}) & (0.0028) & (\textbf{0.0000}) & (\textbf{0.0000}) & (\textbf{0.0000})\\*
	\cmidrule{2-7}
	& \multirow{2}{*}{$1000$} & 0.0000 & 0.0384 & 0.0003 & 0.0003 & 0.0002\\*
	&  & (\textbf{0.0000}) & (0.0025) & (\textbf{0.0000}) & (\textbf{0.0000}) & (\textbf{0.0000})\\
	\midrule
	\multirow{6}{*}{GTTA} & \multirow{2}{*}{$10$} & 0.0000 & 0.0077 & 0.0001 & 0.0001 & 0.0000\\*
	&  & (\textbf{0.0000}) & (\textbf{0.0000}) & (\textbf{0.0000}) & (\textbf{0.0000}) & (\textbf{0.0000})\\*
	\cmidrule{2-7}
	& \multirow{2}{*}{$100$} & 0.0000 & 0.0154 & 0.0001 & 0.0001 & 0.0001\\*
	&  & (\textbf{0.0000}) & (\textbf{0.0000}) & (\textbf{0.0000}) & (\textbf{0.0000}) & (\textbf{0.0000})\\*
	\cmidrule{2-7}
	& \multirow{2}{*}{$1000$} & 0.0000 & 0.0231 & 0.0002 & 0.0002 & 0.0002\\*
	&  & (\textbf{0.0000}) & (\textbf{0.0000}) & (\textbf{0.0000}) & (\textbf{0.0000}) & (\textbf{0.0000})\\
	\midrule
	\multirow{6}{*}{LG} & \multirow{2}{*}{$10$} & \textbf{0.0000} & 0.0006 & 0.0000 & 0.0000 & \textbf{0.0000}\\*
	&  & (\textbf{0.0000}) & (0.0003) & (0.0000) & (0.0000) & (\textbf{0.0000})\\*
	\cmidrule{2-7}
	& \multirow{2}{*}{$100$} & \textbf{0.0000} & 0.0019 & 0.0001 & 0.0001 & \textbf{0.0000}\\*
	&  & (\textbf{0.0000}) & (0.0010) & (0.0000) & (0.0000) & (\textbf{0.0000})\\*
	\cmidrule{2-7}
	& \multirow{2}{*}{$1000$} & \textbf{0.0000} & 0.0046 & 0.0001 & 0.0002 & 0.0000\\*
	&  & (\textbf{0.0000}) & (0.0018) & (0.0000) & (0.0000) & (\textbf{0.0000})\\
	\midrule
	\multirow{6}{*}{MEG} & \multirow{2}{*}{$10$} & 0.0000 & 0.0137 & 0.0001 & 0.0001 & 0.0000\\*
	&  & (\textbf{0.0000}) & (\textbf{0.0000}) & (\textbf{0.0000}) & (\textbf{0.0000}) & (\textbf{0.0000})\\*
	\cmidrule{2-7}
	& \multirow{2}{*}{$100$} & 0.0000 & 0.0240 & 0.0002 & 0.0003 & 0.0000\\*
	&  & (\textbf{0.0000}) & (0.0023) & (\textbf{0.0000}) & (\textbf{0.0000}) & (\textbf{0.0000})\\*
	\cmidrule{2-7}
	& \multirow{2}{*}{$1000$} & 0.0000 & 0.0243 & 0.0004 & 0.0004 & 0.0001\\*
	&  & (\textbf{0.0000}) & (0.0033) & (\textbf{0.0000}) & (\textbf{0.0000}) & (\textbf{0.0000})\\
	\midrule
	\multirow{6}{*}{TD} & \multirow{2}{*}{$10$} & 0.0000 & 0.0054 & 0.0000 & 0.0000 & 0.0000\\*
	&  & (\textbf{0.0000}) & (\textbf{0.0000}) & (\textbf{0.0000}) & (\textbf{0.0000}) & (\textbf{0.0000})\\*
	\cmidrule{2-7}
	& \multirow{2}{*}{$100$} & 0.0000 & 0.0080 & 0.0001 & 0.0001 & 0.0000\\*
	&  & (\textbf{0.0000}) & (0.0000) & (\textbf{0.0000}) & (\textbf{0.0000}) & (\textbf{0.0000})\\*
	\cmidrule{2-7}
	& \multirow{2}{*}{$1000$} & 0.0000 & 0.0106 & 0.0001 & 0.0001 & 0.0000\\*
	&  & (\textbf{0.0000}) & (0.0000) & (\textbf{0.0000}) & (\textbf{0.0000}) & (\textbf{0.0000})\\
	\midrule
	\multirow{6}{*}{WOA} & \multirow{2}{*}{$10$} & 0.0000 & 0.0059 & 0.0000 & 0.0000 & 0.0000\\*
	&  & (\textbf{0.0000}) & (\textbf{0.0000}) & (\textbf{0.0000}) & (\textbf{0.0000}) & (\textbf{0.0000})\\*
	\cmidrule{2-7}
	& \multirow{2}{*}{$100$} & \textbf{0.0000} & 0.0084 & 0.0001 & 0.0001 & 0.0000\\*
	&  & (\textbf{0.0000}) & (0.0001) & (\textbf{0.0000}) & (\textbf{0.0000}) & (\textbf{0.0000})\\*
	\cmidrule{2-7}
	& \multirow{2}{*}{$1000$} & \textbf{0.0000} & 0.0113 & 0.0001 & 0.0001 & 0.0000\\*
	&  & (\textbf{0.0000}) & (0.0045) & (\textbf{0.0000}) & (\textbf{0.0000}) & (\textbf{0.0000})\\
	
\end{longtable}

\begin{longtable}{ccccccc}
\caption{Approximate well-supported approximation results for Learning Dynamics on GAMUT. Each datum in the table represents the average well-supported approximation of experiments for the average policy on randomly generated games. The data in the parentheses is average well-supported approximation for the last iteration.}\\
	\toprule
	\multicolumn{2}{c}{Scenario} & Fictitious Play & Hedge & MWU-exp & MWU-linear & Regret Matching\\
	\midrule
	\midrule
	\endfirsthead\multicolumn{7}{l}{\tablename\ \thetable\ -- (\textit{Continued})} \\
	\toprule
	\multicolumn{2}{c}{Scenario} & Fictitious Play & Hedge & MWU-exp & MWU-linear & Regret Matching\\
	\midrule
	\midrule
	\endhead
	\bottomrule
	\multicolumn{7}{r}{\textit{Continued on next page}} \\
	\endfoot
	\bottomrule
	\endlastfoot
	\multirow{6}{*}{BO} & \multirow{2}{*}{$10$} & 0.3730 & 0.3161 & 0.3361 & 0.3391 & 0.3741\\*
	&  & (0.0051) & (0.0087) & (0.0000) & (0.0000) & (\textbf{0.0000})\\*
	\cmidrule{2-7}
	& \multirow{2}{*}{$100$} & 0.2730 & 0.1258 & 0.1704 & 0.1608 & 0.2019\\*
	&  & (\textbf{0.0000}) & (0.0111) & (0.0000) & (0.0000) & (\textbf{0.0000})\\*
	\cmidrule{2-7}
	& \multirow{2}{*}{$1000$} & 0.1284 & 0.1525 & 0.0997 & 0.1002 & 0.1040\\*
	&  & (\textbf{0.0000}) & (0.0089) & (0.0001) & (0.0001) & (0.0000)\\
	\midrule
	\multirow{6}{*}{CD} & \multirow{2}{*}{$10$} & 0.2374 & 0.2861 & 0.2937 & 0.2937 & 0.2938\\*
	&  & (\textbf{0.0000}) & (0.0153) & (0.0001) & (0.0001) & (\textbf{0.0000})\\*
	\cmidrule{2-7}
	& \multirow{2}{*}{$100$} & 0.3906 & 0.3198 & 0.3976 & 0.3975 & 0.3981\\*
	&  & (\textbf{0.0000}) & (0.0106) & (0.0000) & (0.0000) & (\textbf{0.0000})\\*
	\cmidrule{2-7}
	& \multirow{2}{*}{$1000$} & 0.1000 & 0.0006 & 0.0990 & 0.0992 & 0.1000\\*
	&  & (\textbf{0.0000}) & (0.0012) & (\textbf{0.0000}) & (\textbf{0.0000}) & (\textbf{0.0000})\\
	\midrule
	\multirow{6}{*}{CG} & \multirow{2}{*}{$10$} & 0.4425 & 0.5386 & 0.6196 & 0.6284 & 0.6373\\*
	&  & (0.0317) & (0.0817) & (\textbf{0.0268}) & (0.0414) & (0.0291)\\*
	\cmidrule{2-7}
	& \multirow{2}{*}{$100$} & 0.4432 & 0.3918 & 0.5921 & 0.5540 & 0.6003\\*
	&  & (\textbf{0.0459}) & (0.1708) & (0.0756) & (0.1156) & (0.0599)\\*
	\cmidrule{2-7}
	& \multirow{2}{*}{$1000$} & 0.3694 & \textbf{0.0279} & 0.5385 & 0.4865 & 0.5565\\*
	&  & (0.1125) & (0.0591) & (0.1097) & (0.1638) & (0.0816)\\
	\midrule
	\multirow{6}{*}{GTD} & \multirow{2}{*}{$10$} & 0.5330 & 0.6037 & 0.7613 & 0.7813 & 0.7895\\*
	&  & (0.0164) & (0.0192) & (0.0000) & (0.0000) & (\textbf{0.0000})\\*
	\cmidrule{2-7}
	& \multirow{2}{*}{$100$} & 0.4626 & 0.5047 & 0.6842 & 0.6954 & 0.7259\\*
	&  & (\textbf{0.0000}) & (0.0288) & (0.0000) & (0.0000) & (\textbf{0.0000})\\*
	\cmidrule{2-7}
	& \multirow{2}{*}{$1000$} & 0.4701 & 0.3979 & 0.6641 & 0.6218 & 0.6604\\*
	&  & (\textbf{0.0000}) & (0.0783) & (0.0001) & (0.0001) & (\textbf{0.0000})\\
	\midrule
	\multirow{6}{*}{GTTA} & \multirow{2}{*}{$10$} & 0.5000 & 0.5074 & 0.5001 & 0.5001 & 0.5000\\*
	&  & (\textbf{0.0000}) & (\textbf{0.0000}) & (\textbf{0.0000}) & (\textbf{0.0000}) & (\textbf{0.0000})\\*
	\cmidrule{2-7}
	& \multirow{2}{*}{$100$} & 0.5000 & 0.5153 & 0.5001 & 0.5001 & 0.5001\\*
	&  & (\textbf{0.0000}) & (\textbf{0.0000}) & (\textbf{0.0000}) & (\textbf{0.0000}) & (\textbf{0.0000})\\*
	\cmidrule{2-7}
	& \multirow{2}{*}{$1000$} & 0.5000 & 0.5231 & 0.5002 & 0.5002 & 0.5002\\*
	&  & (\textbf{0.0000}) & (\textbf{0.0000}) & (\textbf{0.0000}) & (\textbf{0.0000}) & (\textbf{0.0000})\\
	\midrule
	\multirow{6}{*}{LG} & \multirow{2}{*}{$10$} & 0.0042 & 0.0059 & 0.0060 & 0.0060 & 0.0060\\*
	&  & (\textbf{0.0000}) & (0.0028) & (0.0002) & (0.0002) & (\textbf{0.0000})\\*
	\cmidrule{2-7}
	& \multirow{2}{*}{$100$} & 0.0047 & 0.0474 & 0.0540 & 0.0540 & 0.0540\\*
	&  & (\textbf{0.0000}) & (0.0076) & (0.0004) & (0.0005) & (\textbf{0.0000})\\*
	\cmidrule{2-7}
	& \multirow{2}{*}{$1000$} & 0.0813 & 0.0827 & 0.0835 & 0.0835 & 0.0835\\*
	&  & (\textbf{0.0000}) & (0.0128) & (0.0005) & (0.0005) & (\textbf{0.0000})\\
	\midrule
	\multirow{6}{*}{MEG} & \multirow{2}{*}{$10$} & 0.6194 & 0.9858 & 0.9999 & 0.9999 & 1.0000\\*
	&  & (\textbf{0.0000}) & (\textbf{0.0000}) & (\textbf{0.0000}) & (\textbf{0.0000}) & (\textbf{0.0000})\\*
	\cmidrule{2-7}
	& \multirow{2}{*}{$100$} & 0.7030 & 0.9693 & 0.9998 & 0.9997 & 1.0000\\*
	&  & (\textbf{0.0000}) & (0.0880) & (\textbf{0.0000}) & (\textbf{0.0000}) & (\textbf{0.0000})\\*
	\cmidrule{2-7}
	& \multirow{2}{*}{$1000$} & 0.5456 & 0.9649 & 0.9996 & 0.9996 & 0.9999\\*
	&  & (\textbf{0.0000}) & (0.0848) & (\textbf{0.0000}) & (\textbf{0.0000}) & (\textbf{0.0000})\\
	\midrule
	\multirow{6}{*}{TD} & \multirow{2}{*}{$10$} & 0.4302 & 0.5882 & 0.5779 & 0.5779 & 0.5779\\*
	&  & (\textbf{0.0000}) & (0.0642) & (\textbf{0.0000}) & (\textbf{0.0000}) & (\textbf{0.0000})\\*
	\cmidrule{2-7}
	& \multirow{2}{*}{$100$} & 0.3869 & 0.6127 & 0.5783 & 0.5782 & 0.5781\\*
	&  & (\textbf{0.0000}) & (0.0504) & (\textbf{0.0000}) & (\textbf{0.0000}) & (\textbf{0.0000})\\*
	\cmidrule{2-7}
	& \multirow{2}{*}{$1000$} & 0.4165 & 0.6345 & 0.5804 & 0.5817 & 0.5836\\*
	&  & (\textbf{0.0000}) & (0.0383) & (\textbf{0.0000}) & (\textbf{0.0000}) & (\textbf{0.0000})\\
	\midrule
	\multirow{6}{*}{WOA} & \multirow{2}{*}{$10$} & 0.2421 & 0.2633 & 0.2514 & 0.2514 & 0.2514\\*
	&  & (\textbf{0.0000}) & (\textbf{0.0000}) & (\textbf{0.0000}) & (\textbf{0.0000}) & (\textbf{0.0000})\\*
	\cmidrule{2-7}
	& \multirow{2}{*}{$100$} & 0.0389 & 0.0723 & 0.0392 & 0.0392 & 0.0390\\*
	&  & (\textbf{0.0000}) & (0.0399) & (\textbf{0.0000}) & (\textbf{0.0000}) & (\textbf{0.0000})\\*
	\cmidrule{2-7}
	& \multirow{2}{*}{$1000$} & 0.0046 & 0.0550 & 0.0050 & 0.0053 & 0.0048\\*
	&  & (\textbf{0.0000}) & (0.0171) & (\textbf{0.0000}) & (\textbf{0.0000}) & (\textbf{0.0000})\\
	
\end{longtable}

\end{document}